\documentclass[%
reprint,
amsmath,amssymb,
 aps,
]{revtex4-2}
\usepackage{etoolbox}
\usepackage{graphicx}
\usepackage{dcolumn}
\usepackage{bm}
\usepackage{svg}
\usepackage{booktabs}
\usepackage{multirow}
\usepackage{longtable} 
\usepackage{titlesec}
\usepackage{graphicx}
\titleformat{\section}[hang]{\normalfont\bfseries}{\thesection}{12em}{}
\usepackage{colortbl}
\usepackage{etoolbox}
\makeatletter
\patchcmd{\@makecaption}{\@ifdim{\wd\@tempboxa >\hsize}}{\@firstoftwo}{}{}
\makeatother

\begin{document}


\title{Strain fingerprinting of exciton valley character}

\author{Abhijeet Kumar$^{1}$}
\thanks{A. Kumar and D. Yagodkin contributed equally}
\author{Denis Yagodkin$^{1}$}
\thanks{A. Kumar and D. Yagodkin contributed equally}
\author{Roberto Rosati$^{2}$}%

\author{Douglas J. Bock$^{1}$}
\author{Christoph Schattauer$^{3}$}
\author{Sarah Tobisch$^{3}$}
\author{Joakim Hagel$^{4}$}
\author{Bianca Höfer$^{1}$}
\author{Jan N. Kirchhof$^{1}$}
\author{Pablo Hernández López$^{5}$}
\author{Kenneth Burfeindt$^{1}$}
\author{Sebastian Heeg$^{5}$}
\author{Cornelius Gahl$^{1}$}
\author{Florian Libisch$^{3}$}
\author{Ermin Malic$^{2}$}
\author{Kirill I. Bolotin$^1$}
\email{kirill.bolotin@fu-berlin.de}
\affiliation{$^1$Department of Physics, Freie Universität Berlin,
Arnimallee 14, 14195 Berlin, Germany}%
\affiliation{$^2$Philipps-Universität Marburg, 35032 Marburg, Germany}
\affiliation{$^3$Institute for Theoretical Physics, TU Wien, Wiedner
Hauptstraße 8-10, 1040 Vienna, Austria}
\affiliation{$^4$Department of Physics, Chalmers University of Technology, 41296 Gothenburg, Sweden}
\affiliation{$^5$Institute for Physics and IRIS Adlershof, Humboldt-Universität Berlin,
Newtonstraße 15, 12489 Berlin, Germany}




\date{\today}
\begin{abstract}
\textbf{Abstract:} Momentum-indirect excitons composed of electrons and holes 
in different valleys define optoelectronic
properties of many semiconductors, but are challenging to detect due
to their weak coupling to light. The identification of an excitons'
valley character is further limited by complexities associated with
momentum-selective probes. Here, we study the photoluminescence of indirect excitons in
controllably strained prototypical 2D semiconductors (WSe$_2$, WS$_2$)
at cryogenic temperatures. We find that
these excitons i)
exhibit valley-specific energy shifts, enabling their valley
fingerprinting, and ii) hybridize with bright excitons, becoming
directly accessible to optical spectroscopy methods. This approach
allows us to identify multiple previously inaccessible excitons with
wavefunctions residing in K, $\Gamma$, or Q valleys in the momentum space as
well as various types of defect-related excitons. Overall, our approach
is well-suited to unravel and tune intervalley excitons in various semiconductors.

\end{abstract}

\maketitle



\begin{center}
    {\textbf{Introduction}}
\end{center}

Two-dimensional semiconductors from the family of transition metal
dichalcogenides (TMDs) host a wide variety of excitons, Coulomb-bound
electron-hole pairs. The electron/hole wavefunctions for the most
relevant excitons reside at the extrema of conduction (CB) and valence
bands (VB) at the K, K', Q, and $\Gamma$ points of the Brillouin
zone\cite{1,2}. Much of the early research was devoted to
momentum-direct excitons with electron and hole wavefunctions in K
valley (KK neutral excitons, charged excitons, biexcitons, etc.). These
quasiparticles with large oscillator strength directly couple to
light and exhibit prominent features in the optical absorption and
emission spectra\cite{1,3,4,5}. However, these "bright" species are only a small subset of excitons in TMDs. Recent studies
show that the optoelectronic properties of TMDs are largely defined by a
much more diverse class of "dark" momentum-indirect intervalley excitons
X$_{\text{AB}}$, quasiparticles with hole and electron wavefunctions
residing in the valleys A (= K, $\Gamma$) and B (= K', Q),
respectively\cite{6,7,8}. Some of these quasiparticles, e.g.,
X$_{\text{KQ}}$, constitute the excitonic ground state of several
TMDs, e.g., monolayers of WSe$_2$, WS$_2$, and bilayers\cite{2,9}.
These states weakly interact with light,
show several orders of magnitude longer
lifetimes, and exhibit long diffusion lengths compared to their direct
counterparts\cite{6,10}. Therefore, the long-range spatial
transport and temporal dynamics observed in these materials are defined
by the interactions and interconversion between direct and indirect
excitons\cite{6,11}. The long lifetime of indirect excitons makes
them a prime candidate for the realization of many-body correlated
states in TMDs\cite{12,13}. Defect-related excitons
constitute another class of quasiparticles with distinct momentum
character. The hole wavefunction for a defect exciton resides in the
K-valley of the VB, while the electron wavefunction (typically localized
at a point defect) is represented by the momentum-delocalized defect states
(D) near the CB\cite{14,15,16}. These defect-related
excitons critically contribute to the long spin/valley lifetime and
carrier dynamics in TMDs\cite{14}.

Despite multiple studies suggesting the defining role of
momentum-indirect excitons in the optical and transport properties of TMDs,
they are much less studied compared to their direct counterparts. The
key challenge is that the conventional optical spectroscopy approaches
cannot directly differentiate between the excitons residing in different
valleys as they are hampered by the lack of momentum resolution and the low
oscillator strength of these quasiparticles\cite{17}.
Additionally, some indirect excitons such as X$_{\text{KK'}}$ or
X$_{\text{KQ}}$ lie in energetic proximity of a series of
tightly-bound KK excitons (trions, biexcitons, etc.) further
complicating their study. In principle, techniques such as
angle-resolved photoemission spectroscopy (ARPES) or electron energy
loss spectroscopy (EELS) provide the momentum resolution needed to
identify the momentum-indirect states\cite{18,19,20,21,22}. However, the typical separation between the excitonic species ($\sim$10~meV) in TMDs is much smaller compared to the energy resolution of these techniques. These approaches also require large-area devices and
are challenging to integrate with external stimuli such as electric and
magnetic fields.

\begin{figure*}
    \includegraphics[width=1\linewidth]{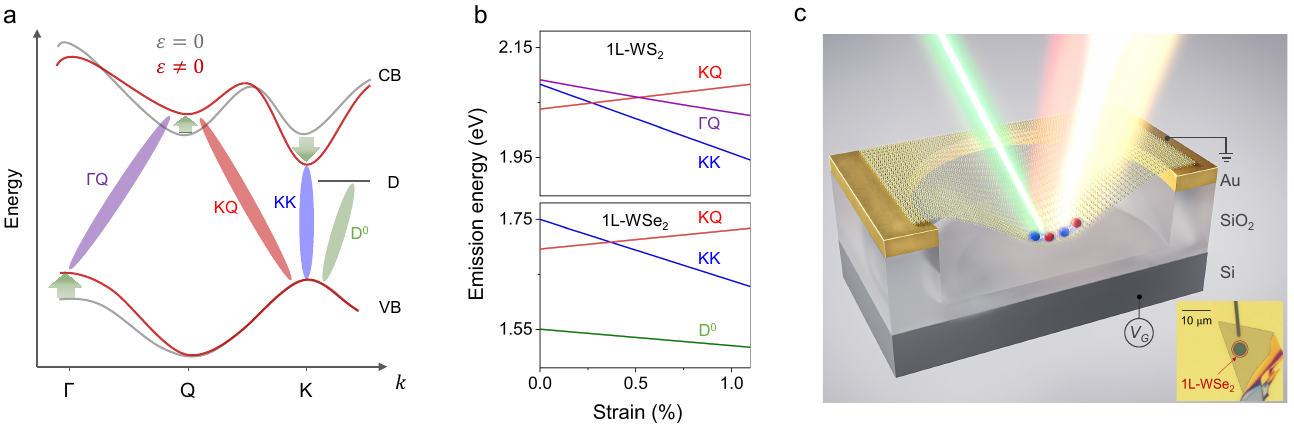}    
    \caption{\textbf{Influence of strain on excitons in TMDs and experimental technique. a)}
    Schematic band structure of 1L-WSe$_2$ under zero (grey)
    and tensile (red) strain. Different valleys respond to strain
    differently (green arrows). 
    \textbf{b)} Calculated energies of various
    intra- and inter-valley excitons vs. biaxial strain in
    1L-WS$_2$ and 1L-WSe$_2$. 
    \textbf{c)}
    Straining technique: an applied gate voltage (\(V_{G}\)) induces biaxial strain \(\varepsilon\) in the center of a suspended TMD monolayer via
    electrostatic forces. Inset shows an optical image of a monolayer
    WSe$_2$ suspended over a circular trench.}

    \label{fig:1}
\end{figure*}

In this work, we identify the valley character of momentum-indirect
excitons in TMDs. We tackle the challenges outlined above by studying
the response of excitonic photoluminescence (PL) peaks to mechanical strain
at cryogenic temperatures. We show that the exciton's energy changes
with strain at a rate characteristic of the valleys in which its
electron and hole wavefunctions reside, allowing straightforward
identification of the exciton valley character --- "valley
fingerprinting". Moreover, a distinct strain response implies that at a
specific strain value, the indirect excitons can be brought into
energetic resonance with the momentum-direct bright excitons. The
hybridization between these species brightens the dark states allowing
their optical detection. We use our technique to directly observe and identify
previously hypothesized, but experimentally inaccessible intervalley KQ
and $\Gamma$Q excitons. In addition, we fingerprint the fine structure of
well-known KK excitons and several types of defect-related excitons. Finally, we achieve \emph{in-situ} strain tuning of
quantum-confined excitons associated with single photon emitters by up
to $80 $~meV. In a larger context, our work establishes strain engineering
as a tool to identify the valley character of excitons, to tune them,
and to explore inter-excitonic interactions.

\begin{center}
    {\textbf{Results}}
\end{center}

\begin{figure*}
    \includegraphics[width=0.9\linewidth]{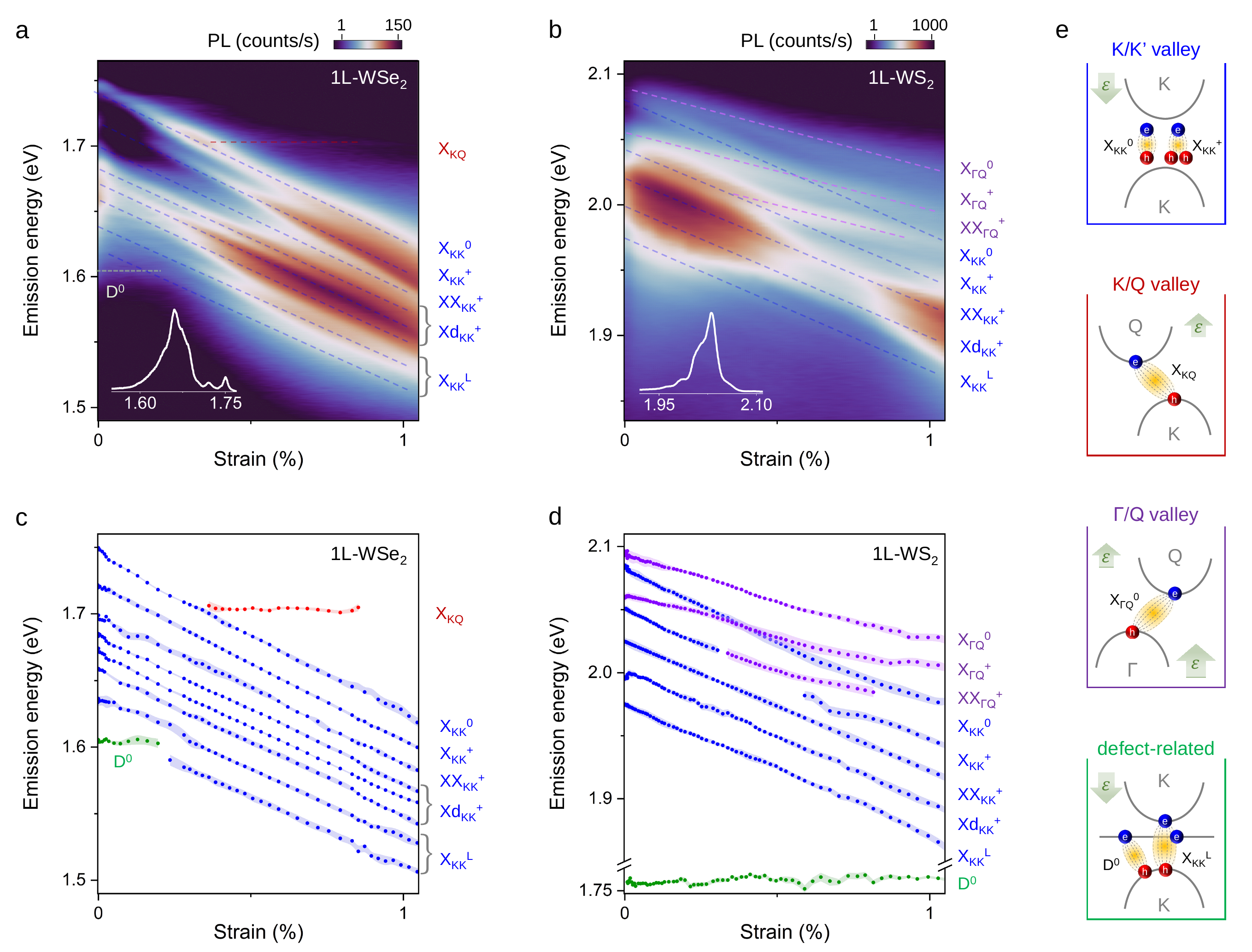}
    \caption{\textbf{Strain response of excitons and identification of
    their valley character. a,b)} False color map of PL vs. strain (log-scale) in
    1L-WSe$_2$ and 1L-WS$_2$ devices at 10~K.
    Dashed lines highlight strain-dependent excitonic peaks: neutral
    excitons (X$^{\text{0}}$), trions (X$^{\text{+}}$), dark
    trions and their phonon replicas (Xd$^{\text{+}}$), defect excitons
    (D$^{\text{0}}$), and excitons bound to defects
    (X$^{\text{L}}_{\text{KK}}$). Insets show the PL spectra
    from devices at \(\varepsilon = 0\). 
    \textbf{c,d)} Extracted peak
    positions vs. strain of various excitons from Fig. 2a and 2b, respectively. The shadows denote uncertainties in the exciton peak positions.
    Groups of peaks corresponding to a specific strain gauge factor are
    color-coded. The analysis of gauge factors allows valley fingerprinting
    of the corresponding excitons. 
    \textbf{e)} Cartoons depicting valley
    compositions of KK (blue), KQ (red), $\Gamma$Q (purple) and defect-related
    (green) excitons. Green arrows indicate a strain-induced shift of the K, Q, and $\Gamma$ valleys
    with respect to the K valley in the VB.}
\end{figure*}

\textbf{Theoretical predictions for excitonic strain response.} 
We start by analyzing the strain response of various intra-
and inter-valley excitons in TMDs theoretically. First-principles calculations
show\cite{2,23,24,25} that the CB minima (\(E^{\text{CB}}\)) and the
VB maxima (\(E^{\text{VB}}\)) in different valleys (e.g., K, Q, $\Gamma$) as well as
the defect state D react differently to an applied strain,
\(\varepsilon\) (tensile biaxial strain unless stated otherwise). For example, the CB at the K valley shifts down in
energy relative to the K valley VB with increasing tensile strain, the
energy of the defect states remains nearly strain-independent, and the
CB at the Q valley shifts up in energy (Fig.~1a). The behaviour of each
valley is governed by the distinct strain dependence of the overlap of
electronic orbitals constituting its wavefunctions, Fig.~S1\cite{23}. In
general, the energy of an exciton X$_{\text{AB}}$ is given by
$E_\text{AB}^{X} = E_\text{B}^{\text{CB}} – E_\text{A}^{\text{VB}} – E_\text{AB}^{bind}$, with
the last term representing the exciton binding energy. Since the binding
energy is only weakly strain-dependent, especially in K and Q valleys
(Fig.~S1), the strain response of an
exciton is predominantly defined by its constituent
valleys\cite{25,26}. The calculated strain response of
various excitons in 1L-WSe$_2$ and 1L-WS$_2$
supports this intuition (Fig.~1b; see Notes~S1 and S2 for calculation
details). For example, K/K'-valley excitons in 1L-WSe$_2$
are predicted to redshift with a strain gauge factor
\({\Omega}_\text{KK} = dE_{{\text{KK}}}^{X}/d\varepsilon =111\)~meV/\%. The $\Gamma$Q
excitons also redshift in 1L-WS$_2$, although at a lower rate 
\({\Omega}_\text{$\Gamma$Q} = 60\)~meV/\%. A $\Gamma$Q exciton in
1L-WSe$_2$ is expected to lie $\sim$~270 meV above
the KK exciton and therefore is not considered in our
study, Fig.~S1\cite{2}. The KQ excitons in 1L-WSe$_2$ shift in energy under
tensile strain with \({\Omega}_\text{KQ}= -34\)~meV/\% while a localized
defect-related exciton D$^{\text{0}}$ is only weakly strain-dependent,
\({\Omega}_\text{D$0$} = 10-30\)~meV/\%.

The calculations in Fig.~1b reveal the following key traits. First, from
the strain response of an exciton alone one can, in principle, determine
its valley character. Second, because different intervalley excitons
shift in energy with strain at different rates, we can control the
energy separation and hence the coupling between them. This, in turn,
enables novel interactions such as brightening of an otherwise dark
X$_{\text{KQ}}$ or D$^{\text{0}}$ exciton via hybridization
with a bright X$_{\text{KK}}$ exciton\cite{6,14,27}.
While these ideas have already been applied to examine the strain response and
hybridization between intervalley excitons,\cite{6,28,29,30,31,32,33,34}
most strain-engineering techniques function only at room
temperature. Under these conditions, temperature-related exciton 
linewidth broadening and
thermal dissociation severely limit the range of accessible excitonic
species. Examining the full range of intervalley excitons and studying
their interactions therefore requires low-disorder TMD devices with \emph{in-situ}
strain control up to a few percent at cryogenic temperature.

\textbf{Experimental realization of exciton valley fingerprinting.}
To address these requirements, we use the electrostatic straining
technique we recently developed\cite{27} (Fig.~1c). A
monolayer flake is suspended over a circular trench in a
Au/SiO$_2$/Si substrate where an applied gate voltage
($V_G$) induces biaxial strain in the center of the
membrane via electrostatic forces. The device is placed inside a
cryostat ($T =  10$~K), while its PL response to an optical
excitation is measured as a function of strain in
the center of the suspended flake (see \emph{Methods}). The induced strain is symmetric with
respect to the $V_G$ polarity (Fig.~S2) and reaches $1.5\%$, limited by the dielectric breakdown of SiO$_{2}$.

Figure 2a and 2b show photoluminescence emission spectra vs. strain
of 1L-WSe$_2$ and 1L-WS$_2$, respectively
(see Note~S4, Fig.~S3 for strain determination). Both material systems
exhibit complex spectra with an abundance of excitonic peaks (insets of
Fig.~2a,b). Some well-known excitons such as neutral and charged
excitons (X$_\text{KK}^0$,
X$_\text{KK}^+$), charged biexcitons
(XX$_\text{KK}^+$), dark trions
(Xd$_\text{KK}^+$) and their phonon replicas can
be identified at zero strain by comparing their peak
positions and power dependence with previous reports\cite{35,rosati2020temporal,rosati2020strain}
(Fig.~S4, S5). Once the strain is applied, different groups of peaks
exhibit distinct strain dependence. We extract the energy positions of
various excitonic peaks vs. strain and color-code each group of peaks
based on their strain dependence (Fig.~2c,d). We now proceed to
assign these peaks to the excitons residing in specific valleys of the electronic band structure.

\begin{figure*}
    \includegraphics[width=0.91\linewidth]{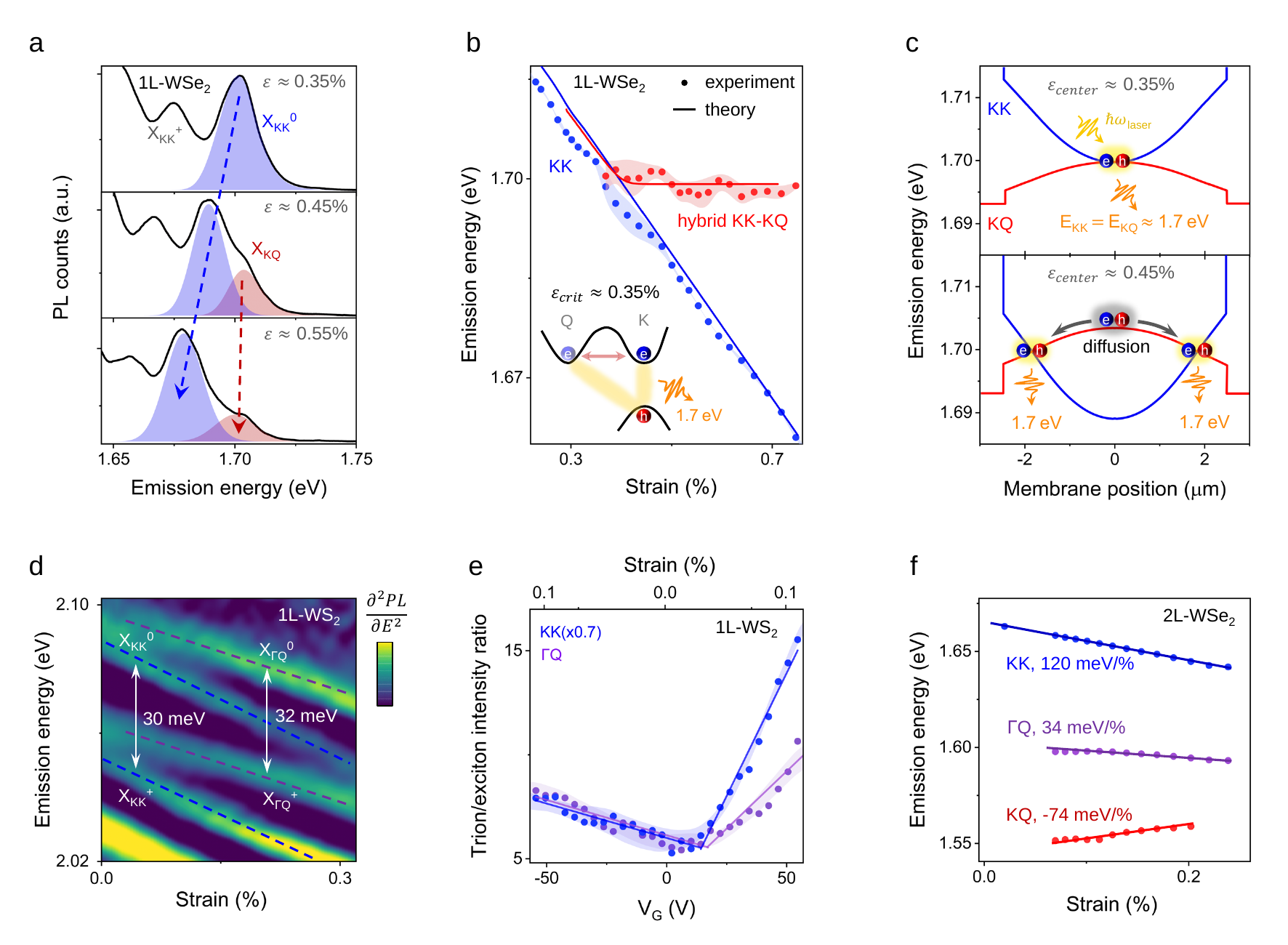} 
    \caption{\textbf{Momentum-indirect Q-valley excitons in WSe$_2$ and WS$_2$. a)} PL spectra 
    in 1L-WSe$_2$ (device 2, different from Fig. 2) at selected strain values.
    Above a critical strain value \(\varepsilon_{crit}\ge0.35\%\), a
    strain-independent feature (X$_\text{KQ}$)
    emerges on the blue side of X$_\text{KK}^0$. The
    two peaks are highlighted by Gaussian fits. \textbf{b)}
    Extracted emission energy of X$_\text{KK}^0$
    (blue) and X$_\text{KQ}$ (red) in
    1L-WSe$_2$ vs. strain; the shadows denote the uncertainty bar. The
    solid lines are the theoretically predicted emission energy of the KK
    and the hybrid KK-KQ excitons. The inset depicts the hybridization
    scenario and corresponding emission when the local
    strain is \(\varepsilon_{crit}\approx0.35\)\%. \textbf{c)} Energies of the
    KK and KQ excitons vs. position on a line cut across the membrane for
    two different strain values (characterized in the center of the
    membrane, \(\varepsilon_{center}\)). A strain inhomogeneity
    (\({\Delta}\varepsilon/\varepsilon_{center}\approx 10\%/\mu \text{m}\))
    causes spatially dependent resonance condition
    $E_{\text{KK}} = E_{\text{KQ}}$. A KQ exciton
    diffuses (anti-funnels) and emits light from the position within the membrane where
    the resonance conditions are achieved. \textbf{d)} False color map of
    \(d^{2}PL/dE^{2}\) vs. strain in 1L-WS$_2$ highlighting
    individual excitonic peaks. Blue and purple lines indicate peaks with
    gauge factors corresponding to KK and $\Gamma$Q valley excitons, respectively.
    The energy difference between a pair of KK peaks and a pair of $\Gamma$Q peaks
    corresponds to the binding energy of charged excitons (trions).
    \textbf{e)} Trion to exciton intensity ratio (T/A, calculated from the ratio of
    areas under the corresponding peaks) vs. $V_G$ for KK
    (blue) and $\Gamma$Q (purple) excitons in weak strain regime (\(\varepsilon
    \le 0.2\%\)). For $V_G \ge +20$~V,
    the T/A ratio for KK and $\Gamma$Q excitons increases sharply suggesting that the
    Fermi level has entered the CB. 
    \textbf{f)} Extracted exciton peak positions vs. strain in a
    2L-WSe$_2$ device. Strain gauge factors of these states
    are close to the theoretical expectations for KK, $\Gamma$Q, and KQ excitons in 2L-WSe$_2$. Note that a prestrain may influence the energy positions of the observed excitons\cite{kirchhof2022nanomechanical}.}
\end{figure*}

\textbf{KK valley excitons.} We first focus on the largest group of peaks
(X$_\text{KK}^0$,
X$_\text{KK}^+$, XX$_\text{KK}^+$, Xd$_\text{KK}^+$ ) shifting down in energy with
gauge factor \(\Omega_\text{KK} = 118\pm6\)~meV/\% in 1L-WSe$_2$
and \(\Omega_\text{KK} = 102\pm13\)~meV/\% in 1L-WS$_2$ (blue in
Fig.~2c,d). This is the shift expected for an optical transition
between the VB and CB at the K/K' valley (blue in Fig.~1b), in
agreement with previous reports\cite{36,29,31,32,33}. We find very close
gauge factors for species within the KK group (Fig.~S4). This similarity
suggests that i) effects related to the carrier density changes with
$V_{G}$, estimated to be $< 1.5 \cdot
10^{12}$~cm$^{-2}$ in our
technique\cite{37}, are insignificant compared to the
strain-related effects (Fig.~S4); ii) strain-related changes in the
phonon energies are minor, since the energy spacing between excitons and
their replicas depends on the phonon energy\cite{26,38}; and
iii) effective masses near the K valley are nearly strain-independent
(as suggested by theory, see Fig.~S1) since the binding
energy of the biexciton (determined from the energy difference
between X$_\text{KK}$ and XX$_\text{KK}^-$) is affected by strain only via the effective mass.

\textbf{KQ excitons.} We now turn to the nearly strain-independent
peak at 1.70~eV in 1L-WSe$_2$ (red in Fig.~2c). To assign
the nature of this feature, we plot PL spectra at three applied strain
values of 0.35\%, 0.45\%, and 0.55\% (Fig.~3a). While
X$_\text{KK}^0$ (blue shaded) and
X$_\text{KK}^+$ redshift with strain, a new
feature (red shaded) appears on the high-energy side of
X$_\text{KK}^0$ above a critical strain value,
\(\varepsilon_{crit}\approx 0.35\)\%. This peak exhibits a linear power
dependence suggesting its free excitonic character (Fig.~S5) and emerges exactly at the strain value at which the
X$_\text{KK}^0$ and X$_\text{KQ}$ excitons
are predicted to come into resonance (\(0.35\%\) in Fig.~1b;
$E_\text{KK} = E_\text{KQ}=1.705 $~eV). We therefore
suggest that the 1.70~eV peak corresponds to a hybridized state of KQ and
KK excitons. In this scenario, a normally dark KQ exciton acquires oscillator strength
through resonant hybridization with a bright KK exciton with which it
shares the hole wavefunction (inset, Fig.~3b). The emerging hybrid state
should emit only in the vicinity of \(\varepsilon_{crit}\) while having a vanishing
oscillator strength otherwise.

Notably, we observe that the peak at 1.70~eV persists for the applied strain
exceeding 0.35\% and its energy remains nearly strain-independent afterward (Fig. 3b). We ascribe this behavior to a slight parabolic strain variation
(\({\Delta}\varepsilon/\varepsilon_{center}\approx 10\%/\mu\text{m}\)) in our membrane around the maximum reached in the center (Fig.~3c, Fig.~S6). With increasing $V_G$, the strain
first reaches the critical value in the center of the membrane (Fig.~3c,
top panel). When the applied strain \(\varepsilon_{center}\) exceeds 0.35\%, the
condition  \(\varepsilon_{crit}\approx0.35\%\) required for the KK-KQ hybridization
progressively shifts outward from the center of the membrane in a
donut-shaped region (Fig.~3c, bottom panel). Therefore, the emission at
1.70~eV persists even for \(\varepsilon_{center}>0.35\%\) and remains
strain independent (Fig.~3a,b). Our theoretical many-particle model
for the KK-KQ hybridization accounting for a strain inhomogeneity
(\({\Delta}\varepsilon/\varepsilon_{center}\) approximated by a gaussian distribution of FWHM 6.1 $\mu$\text{m}) and
X$_\text{KQ}$ diffusion\cite{6} confirms the
strain-independent behaviour (solid lines in Fig.~3b; see Note~S3 for
details). Decrease of the
hybrid exciton intensity  
for \(\varepsilon_{center}>\varepsilon_{crit}\) further supports our model (Fig.~S7). We
also observe signatures of a similar peak in 1L-WS$_2$ in
the predicted hybridization regime (Fig.~S8). The apparent brightness of hybrid excitons at
large strain values ($\sim1\%$) may be further influenced by high strain non-uniformity near the membrane
edges---this effect could be reduced by increasing the membrane diameter. We note that the KQ excitons
were previously reported in ARPES measurements\cite{18} and
invoked to explain excitonic transport\cite{6,berghauser2018mapping} but, to the
best of our knowledge, have never been directly seen before via optical
spectroscopy techniques. 

While the strain response of
the KQ exciton in a monolayer is governed by hybridization, its
fingerprinting is more straightforward in a bilayer WSe$_2$ (Fig.~3f). Here, 
the KQ emission is more intense due to an abundance of phonons mitigating excess 
momentum and a significantly lower energy of X$_\text{KQ}$ compared to 
X$_\text{KK}$ ensuring its higher population\cite{9}. Indeed, we observe a
low-intensity peak $\sim$120~meV below
X$_\text{KK}^0$ shifting up in energy as predicted for a 
KQ exciton\cite{41} (Fig.~S1, S9).

\textbf{$\mathbf{\Gamma}$Q excitons.} According to our calculations, the $\Gamma$Q and the KK excitons
in unstrained WS$_2$ are nearly resonant, however, can be
distinguished under strain due to distinct gauge factors. To this end,
we note the group of three peaks in Fig.~2d (purple points) red shifting
with gauge factor \(\Omega_\text{$\Gamma$Q}=63\pm8\)~meV/\%. This gauge factor as well as the
energy of the states match theoretical expectations for $\Gamma$Q excitons
(Fig.~1b). Within this group, we assign the two highest lying states as
neutral (X$_\text{$\Gamma$Q}^0$) and charged $\Gamma$Q excitons
(X$_\text{$\Gamma$Q}^+$), respectively. To support our
assignment, we first note that the two states are separated by
$\sim$32~meV, a value similar to the
X$_\text{KK}^+$ binding energy of 30~meV\cite{39} (Fig.~3d). Second, these states exhibit the
characteristic behavior of the charged states, i.e., an increase of neutral to charged exciton conversion with
increasing carrier density (Fig.~3e). Finally, we suggest that the
lowest-energy $\Gamma$Q state ($\sim$52 meV below
X$_\text{$\Gamma$Q}^0$ in Fig.~2d) is of biexcitonic
nature since it shows a super-linear dependence of PL on the excitation
power (PL \(\propto\) $P^{1.36}$, Fig.~S5). 
Further work is needed to pinpoint the exact configuration and
brightening mechanism of this state\cite{40}. We highlight that our 
suspended devices are ideally suited to study $\Gamma$Q excitons: 
the X$_\text{$\Gamma$Q}$ energy 
in a supported device is affected by screening\cite{klots2018controlled} and is predicted to lie $\sim$50~meV higher with 
significantly weaker emission, rendering their observation challenging (Note~S1). To 
the best of our knowledge, this is the first experimental observation of $\Gamma$Q
excitons in monolayers.

Another system where $\Gamma$Q excitons have been theoretically predicted is
bilayer WSe$_2$ (Fig.~3f). There, the $\Gamma$ and the K valleys
are nearly degenerate in the VB and a $\Gamma$Q exciton is predicted to lie
$\sim$50~meV above their KK counterpart, Fig.~S1). Indeed,
our PL data for bilayer WSe$_2$ indicate a state
$\sim$70~meV below X$_\text{KK}^0$
shifting down in energy with a rate of $\sim34\pm8$~meV/\%, consistent with the
expected gauge factors for $\Gamma$Q excitons in this
material\cite{9} (Fig.~S1,~S9). We note that, unlike in the monolayer case, additional effects such as heterostrain or change in
layer separation may complicate the strain response of bilayers.

\textbf{Localized excitons.} Having identified most of the free
excitons, we now proceed to the emission features of
localized states distinguished by sublinear power dependence (Fig.~S5).
The energy of the peak labeled D$^\text{0}$ in Fig.~2c
(green) is nearly strain-independent (\(\Omega_\text{D$\text{0}$}=8\pm10\)~meV/\%), suggesting that the valleys
hosting the corresponding electron/hole wavefunctions shift with strain
at nearly equivalent rates. Comparison with Fig.~1b allows us to
identify D$^\text{0}$ as a chalcogen-vacancy-related defect exciton, consistent with previous study\cite{27}. This state involves an optical
transition between the VB at the K-point and the momentum-delocalized
defect state below the CB (Fig.~1b). Interestingly, also the peaks
$\sim$120~meV and $\sim$150~meV below
X$_\text{KK}^0$, labelled
X$_\text{KK}^{\text{L}}$, exhibit sublinear power
dependence. These peaks, however, exhibit a gauge factor \(106\pm4\)~meV/\%,
close to that of KK excitons. We therefore attribute the peaks,
X$_\text{KK}^{\text{L}}$, to the recombination
of a neutral KK exciton bound to a defect. Indeed, states in the same
energy range have been recently ascribed to defect-bound
excitons\cite{35}.

\begin{figure}
    \includegraphics[width=1\linewidth]{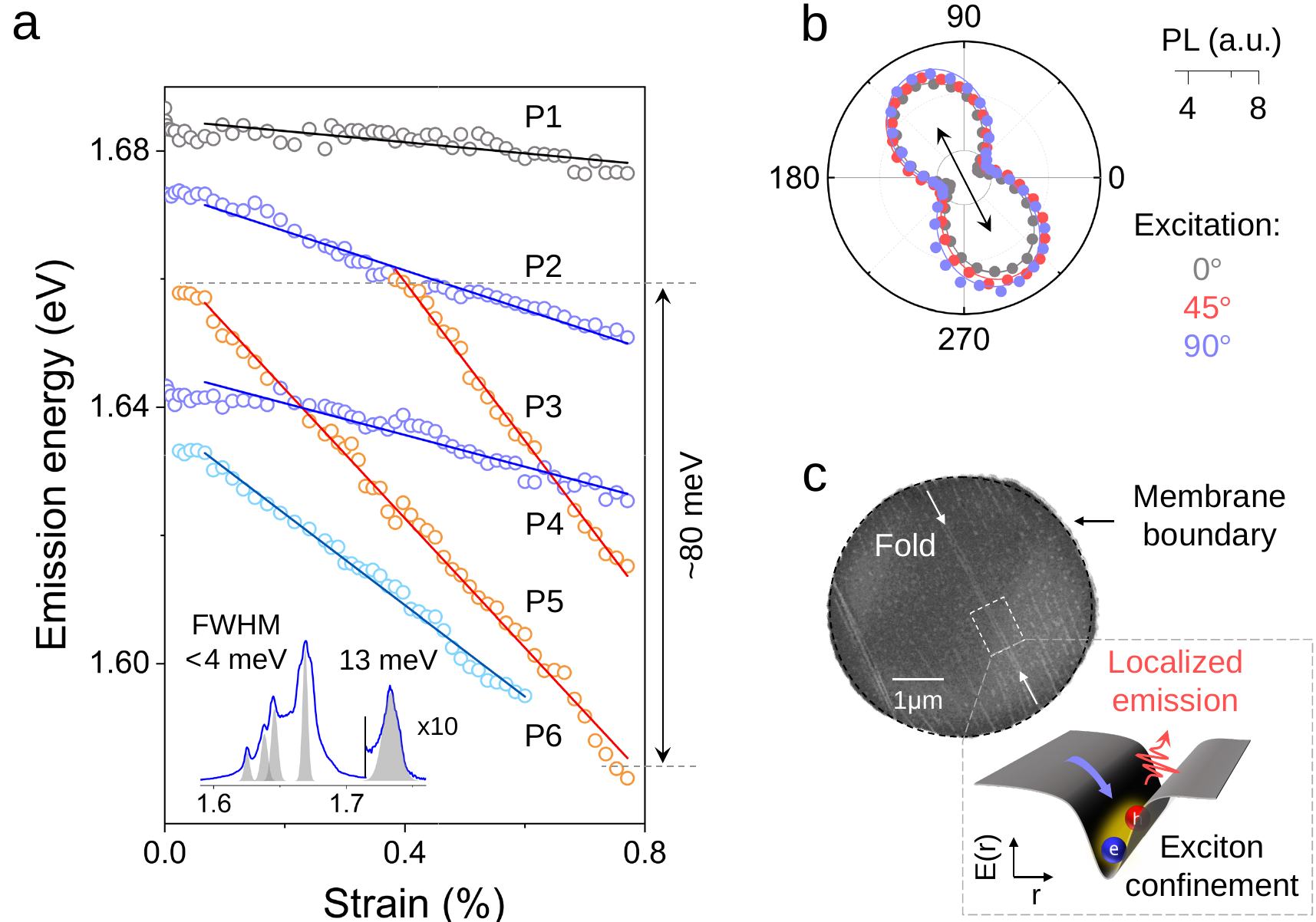}    
    \caption{\textbf{Quantum-confined states in
    1L-WSe$_2$} \textbf{a)} Extracted energy positions vs.
    strain for sharp peaks in 1L-WSe$_2$ (device 3), color-coded
    by gauge factor. The inset shows a PL line cut from the unstrained device. Note
    several sharp features with FWHM up to 4 times narrower compared to
    X$_\text{KK}^{0}$. 
    \textbf{b)}
    Polarization-resolved emission of peak the P4 for three different directions
    of excitation polarization (grey, red, blue). The emission exhibits a
    preferred direction (black arrow), and is independent of the excitation
    polarization. 
    \textbf{c)} Top: SEM image of the membrane reveals
    several sub-\(\mu\)m folds along the direction of polarization of P4;
    Bottom: a cartoon depicting localized emission from an exciton trapped in
    a sharp potential.}
    \label{fig:4}
\end{figure}

A similar scenario is observed in 1L-WS$_2$ (Fig.~2d): the
strain-independent peak D$^{\text{0}}$ appears near
1.75~eV, and the X$_\text{KK}^{\text{L}}$
peak $\sim$1.97~eV (at \(\varepsilon = 0\)). The energy difference
between D$^{\text{0}}$ and X$_\text{KK}^{\text{0}}$
is larger in WS$_2$ compared to WSe$_2$
($\sim$340 vs. $\sim$150~meV), suggesting different
defect energy levels involved in the corresponding optical transitions
(Fig.~S1). We see that, in principle, the strain response allows
distinguishing between various types of defect-related excitons
including two-particle (e.g., D$^{\text{0}}$) or three-particle
(X$_\text{KK}^{\text{L}}$) states.

In addition to D$^\text{0}$ and
X$_\text{KK}^{\text{L}}$ consistently seen in multiple
devices, in some devices we observe additional localized exciton peaks.
Figure~4 shows strain response in one such WSe$_2$ sample
where we identify a group of closely lying peaks labeled P1---P6 with
various gauge factors in the range $10-100$~meV/\%. These peaks are
$\sim$4~times narrower than
X$_\text{KK}^{\text{L}}$ (inset in Fig.~4a), have a
preferred emission direction independent of the excitation (Fig.~4b, Fig.~S10), and show sublinear
power dependence (Fig.~S10). These features are tell-tale signs of
extensively studied\cite{42,43,44,45} single-photon emitters in
WSe$_2$. A locally inhomogeneous strain profile has been
suggested as the critical requirement for the formation of such emitters
via exciton confinement\cite{14,27}. Indeed, the scanning
electron microscope (SEM) image of this device shows the presence of
folds across the membrane (Fig.~4c, Fig.~S10) that could lead to sharp
(up to 0.02~\%/nm) strain gradients\cite{46,47}. Notably, the
PL polarization from peaks P1---P6 is oriented along those folds confirming exciton confinement (Fig.~4b).
The variation in the gauge factors is then likely associated with
the effect of the externally applied strain on the inhomogeneous strain
present in the fold. Crucially, the energy of emitters P1---P6 can be
tuned by up to 80~meV under the application of strain. That is the
largest reported tuning for such states, to the best of our
knowledge\cite{48,49}. Moreover, two distinct states can be brought into an
energetic resonance at specific strain values resulting in enhanced emission (Fig.~S10), possibly
caused by exciton population redistribution\cite{50}. These
observations suggest the potential of confined states with engineered
strain gauge factors to control exciton hybridization. In the
future, $g^2$ measurements are of particular
interest for proving the indistinguishability of such resonant states\cite{patel2010two}.

\begin{center}
    {\textbf{Discussion}}
\end{center}
To summarize, we established mechanical strain as a powerful tool to
brighten the momentum-indirect dark excitons and fingerprint their
valley character. In our approach, \emph{in-situ} control over
mechanical strain at low temperatures is the key to unveiling the complex
excitonic landscape of 1L-WSe$_2$, 1L-WS$_2$,
and 2L-WSe$_2$. We established the valley character of
previously inaccessible $\Gamma$Q excitons/trions, KQ excitons 
in addition to the defect-related excitons (Table 1). We also identified a brightening mechanism for the normally dark KQ excitons via strain-driven hybridization 
with bright excitons. Finally, we showed wide-range strain tuning of the
energy of quantum-confined excitons in 1L-WSe$_2$ opening
pathway to broadly tunable quantum emitters. We note that our current samples have broader excitonic linewidth compared to the 
hBN-encapsulated devices\cite{35,37}. Superior samples with narrower linewidth will enhance the control 
over closely lying excitonic species. Devices with

\setlength{\tabcolsep}{5pt} 
\renewcommand{\arraystretch}{1.05} 
\setlength\LTleft{40pt}    
\setlength\LTright{40pt}           
\setlength\LTcapwidth{2\linewidth}

\begin{longtable*}[b]{@{}rcccclclc|lcccclccclc@{}}
\caption{The measured emission energy at zero applied strain ($E(\varepsilon = 0)$), the strain gauge factor ($\Omega$) and the power law exponent ($\alpha$) for several excitons in WSe$_2$ and WS$_2$. Note that a prestrain may 
influence the $E(\varepsilon = 0)$.
Theoretical results are noted with~*.}\\

\hline
\multicolumn{1}{l}{}                  & \multicolumn{8}{c|}{WSe$_2$}                                                             &  & \multicolumn{10}{c}{WS$_2$}                                                                                   \\ \hline \endfirsthead
\hline \hline
\endlastfoot
\multicolumn{1}{c}{\multirow{2}{*}{}} & \multicolumn{4}{c}{KK}                               &  & KQ           &  & Defect       &  & \multicolumn{4}{c}{KK}                              &  & \multicolumn{3}{c}{$\Gamma$Q}      &  & Defect       \\
\multicolumn{1}{c}{}                  & X$^\text{0}$ & X$^\text{+}$ & XX   & X$^\text{L}$ &  & X$^\text{0}$ &  & D$^\text{0}$ &  & X$^\text{0}$ & X$^\text{+}$ & XX  & X$^\text{L}$ &  & X$^\text{0}$ & X$^\text{+}$ & XX   &  & D$^\text{0}$ \\
$E(\varepsilon = 0)$, eV              & 1.74         & 1.72         & 1.70 & 1.64            &  & 1.69*        &  & 1.60        &  & 2.08         & 2.05         & 2.03 & 1.98            &  & 2.09         & 2.06         & 2.04 &  & 1.76         \\
$\Omega$, meV/\%                      & 118          & 114          & 113  & 103             &  & -34*         &  & 8          &  & 102          & 101          & 100  & 100             &  & 68           & 55           & 56   &  & -6            \\
$\alpha$                              & 0.97            & 1.11            & 1.42    & 0.70               &  & 0.95           &  & 0.66         &  & 0.97          & 0.99          & 1.33  & 0.86            &  & 1.01          & 1.18         & 1.36  &  & 0.50          \\ 
\end{longtable*}
\noindent engineered strain inhomogeneity may allow us to better manipulate the quantum emitters,
spatially modulate exciton hybridization, and guide exciton transport. Extension of the
simulations to account for non-linear effects and inhomogeneous strain might be required 
beyond the linear strain regime analyzed here. 

Our approach to brighten and fingerprint intervalley excitons via strain
engineering opens multiple possibilities for future research. First, it
may be applied to identify the valley character of excitons in other 
systems such as moiré TMD heterostructures\cite{51}, perovskites\cite{52,53}, and 2D magnets\cite{Cenker2022,Diederich2022}. Second, our technique
enables optical control of spins in the Q valley that remained
unexplored until now\cite{54}. The spin/valley locked KQ
excitons should exhibit a long lifetime and diffusion length, hence may prove
advantageous for spin-/valleytronics compared to the direct
excitons\cite{6,10,55}. We expect a pronounced
strain-dependence of the transport dynamics and the lifetimes of these
excitons. Third, our approach may be capable of detecting
strain-dependent changes in the effective masses of
excitons\cite{26}, controlling exciton-phonon interactions
(by modulating the energy separation between the states
involved)\cite{28,37,40}, and distinguishing different types of
defects (since various defect states exhibit distinct strain
response)\cite{14,27}. The application of uniaxial strain may
further break the symmetry of our system thereby changing the band
topology of the corresponding excitons\cite{56}. This enables
unique prospects to manipulate valley pseudospins via large in-plane
pseudo magnetic fields\cite{57,iakovlev2023fermi}.

{\textbf{Methods}}

\textit{Sample fabrication}
The devices were fabricated by dry transfer of mechanically exfoliated
TMD flakes onto a circular trench (diameter is $\sim$5 $\mu$m) wet
etched via Hydrofluoric (HF) acid in Au/Cr/SiO$_2$/Si stack. The
strain in the membrane is induced by applying a gate voltage (typically
in the range of up to $\pm$210~V) between the TMD flake (electrically
grounded) and the Si back gate of the chip. The strain in the center was
characterized using laser interferometry (see Note~S4).

\textit{Optical measurements}
The devices were measured inside a cryostat (CryoVac Konti Micro) at a base temperature of 10~K. The PL measurements were carried
out using the Spectrometer Kymera 193i Spectrograph, while CW lasers with \(\lambda=532\)~nm (10~$\mu$W) and \(\lambda=670\)~nm (6~$\mu$W) tightly focused in the center of the membrane with spot diameter $\sim$1~$\mu$m were used to excite 
WS$_2$ and WSe$_2$, respectively. Polarization-resolved PL measurements were performed using a combination of a half-wave plate (RAC 4.2.10, B. Halle) and an analyzer (GL~10, Thorlabs) before the spectrometer to select specific polarization.
The fold was confirmed using a Scanning Electron Microscopy (SEM) system
Raith Pioneer II SEM/EBL at an accelerating voltage of $10$~kV. 

\textbf{Acknowledgements}

The Berlin group acknowledges the Deutsche
Forschungsgemeinschaft (DFG) for financial support through the
Collaborative Research Center TRR 227 Ultrafast Spin Dynamics (project
B08 and the Federal Ministry of Education and Research
(BMBF, Projekt 05K22KE3). The Marburg group acknowledges financial 
support by the DFG via SFB 1083 (project B9) and the regular
project 512604469. The Vienna group acknowledges financial support by
The Austrian Science Fund (FWF) through doctoral college TU-DX (DOC 142-N) and the DOC-fellowship of the Austrian Academy of Sciences. P.H.L. and S.H. acknowledge funding from the DFG under the Emmy Noether Initiative (project-ID 433878606).

\textbf{Author Contributions}

A.K., D.Y., and K.I.B. conceived the project. A.K., D.Y., and C.G. designed the experimental setup. A.K.,
D.Y., D.J.B., B.H., and K.B. prepared the samples. S.H. and P.H.L. developed the electrostatic straining technique. A.K., D.Y., and
D.J.B. performed the optical measurements. R.R., J.H., and E.M.
developed theory for excitons. C.S., S.T., and F.L. developed the theory for
defect states. J.N.K. performed mechanical simulations. A.K. and D.Y.
analyzed the data. A.K., D.Y., and K.I.B. wrote the manuscript with
input from all co-authors.

\textbf{Data Availability Statement}

The data that support the findings of this study are available from the
corresponding author upon reasonable request.

The authors declare no competing financial interest.

\bibliography{My_Library_V1}

\begin{thebibliography}{66}%
\makeatletter
\providecommand \@ifxundefined [1]{%
 \@ifx{#1\undefined}
}%
\providecommand \@ifnum [1]{%
 \ifnum #1\expandafter \@firstoftwo
 \else \expandafter \@secondoftwo
 \fi
}%
\providecommand \@ifx [1]{%
 \ifx #1\expandafter \@firstoftwo
 \else \expandafter \@secondoftwo
 \fi
}%
\providecommand \natexlab [1]{#1}%
\providecommand \enquote  [1]{``#1''}%
\providecommand \bibnamefont  [1]{#1}%
\providecommand \bibfnamefont [1]{#1}%
\providecommand \citenamefont [1]{#1}%
\providecommand \href@noop [0]{\@secondoftwo}%
\providecommand \href [0]{\begingroup \@sanitize@url \@href}%
\providecommand \@href[1]{\@@startlink{#1}\@@href}%
\providecommand \@@href[1]{\endgroup#1\@@endlink}%
\providecommand \@sanitize@url [0]{\catcode `\\12\catcode `\$12\catcode `\&12\catcode `\#12\catcode `\^12\catcode `\_12\catcode `\%12\relax}%
\providecommand \@@startlink[1]{}%
\providecommand \@@endlink[0]{}%
\providecommand \url  [0]{\begingroup\@sanitize@url \@url }%
\providecommand \@url [1]{\endgroup\@href {#1}{\urlprefix }}%
\providecommand \urlprefix  [0]{URL }%
\providecommand \Eprint [0]{\href }%
\providecommand \doibase [0]{http://dx.doi.org/}%
\providecommand \selectlanguage [0]{\@gobble}%
\providecommand \bibinfo  [0]{\@secondoftwo}%
\providecommand \bibfield  [0]{\@secondoftwo}%
\providecommand \translation [1]{[#1]}%
\providecommand \BibitemOpen [0]{}%
\providecommand \bibitemStop [0]{}%
\providecommand \bibitemNoStop [0]{.\EOS\space}%
\providecommand \EOS [0]{\spacefactor3000\relax}%
\providecommand \BibitemShut  [1]{\csname bibitem#1\endcsname}%
\let\auto@bib@innerbib\@empty
\bibitem [{\citenamefont {Wang}\ \emph {et~al.}(2018)\citenamefont {Wang}, \citenamefont {Chernikov}, \citenamefont {Glazov}, \citenamefont {Heinz}, \citenamefont {Marie}, \citenamefont {Amand},\ and\ \citenamefont {Urbaszek}}]{1}%
  \BibitemOpen
  \bibfield  {author} {\bibinfo {author} {\bibfnamefont {G.}~\bibnamefont {Wang}}, \bibinfo {author} {\bibfnamefont {A.}~\bibnamefont {Chernikov}}, \bibinfo {author} {\bibfnamefont {M.~M.}\ \bibnamefont {Glazov}}, \bibinfo {author} {\bibfnamefont {T.~F.}\ \bibnamefont {Heinz}}, \bibinfo {author} {\bibfnamefont {X.}~\bibnamefont {Marie}}, \bibinfo {author} {\bibfnamefont {T.}~\bibnamefont {Amand}}, \ and\ \bibinfo {author} {\bibfnamefont {B.}~\bibnamefont {Urbaszek}},\ }\href {\doibase 10.1103/RevModPhys.90.021001} {\bibfield  {journal} {\bibinfo  {journal} {Reviews of Modern Physics}\ }\textbf {\bibinfo {volume} {90}},\ \bibinfo {pages} {021001} (\bibinfo {year} {2018})}\BibitemShut {NoStop}%
\bibitem [{\citenamefont {Malic}\ \emph {et~al.}(2018)\citenamefont {Malic}, \citenamefont {Selig}, \citenamefont {Feierabend}, \citenamefont {Brem}, \citenamefont {Christiansen}, \citenamefont {Wendler}, \citenamefont {Knorr},\ and\ \citenamefont {Berghäuser}}]{2}%
  \BibitemOpen
  \bibfield  {author} {\bibinfo {author} {\bibfnamefont {E.}~\bibnamefont {Malic}}, \bibinfo {author} {\bibfnamefont {M.}~\bibnamefont {Selig}}, \bibinfo {author} {\bibfnamefont {M.}~\bibnamefont {Feierabend}}, \bibinfo {author} {\bibfnamefont {S.}~\bibnamefont {Brem}}, \bibinfo {author} {\bibfnamefont {D.}~\bibnamefont {Christiansen}}, \bibinfo {author} {\bibfnamefont {F.}~\bibnamefont {Wendler}}, \bibinfo {author} {\bibfnamefont {A.}~\bibnamefont {Knorr}}, \ and\ \bibinfo {author} {\bibfnamefont {G.}~\bibnamefont {Berghäuser}},\ }\href {\doibase 10.1103/PhysRevMaterials.2.014002} {\bibfield  {journal} {\bibinfo  {journal} {Physical Review Materials}\ }\textbf {\bibinfo {volume} {2}},\ \bibinfo {pages} {014002} (\bibinfo {year} {2018})}\BibitemShut {NoStop}%
\bibitem [{\citenamefont {He}\ \emph {et~al.}(2014)\citenamefont {He}, \citenamefont {Kumar}, \citenamefont {Zhao}, \citenamefont {Wang}, \citenamefont {Mak}, \citenamefont {Zhao},\ and\ \citenamefont {Shan}}]{3}%
  \BibitemOpen
  \bibfield  {author} {\bibinfo {author} {\bibfnamefont {K.}~\bibnamefont {He}}, \bibinfo {author} {\bibfnamefont {N.}~\bibnamefont {Kumar}}, \bibinfo {author} {\bibfnamefont {L.}~\bibnamefont {Zhao}}, \bibinfo {author} {\bibfnamefont {Z.}~\bibnamefont {Wang}}, \bibinfo {author} {\bibfnamefont {K.~F.}\ \bibnamefont {Mak}}, \bibinfo {author} {\bibfnamefont {H.}~\bibnamefont {Zhao}}, \ and\ \bibinfo {author} {\bibfnamefont {J.}~\bibnamefont {Shan}},\ }\href@noop {} {\bibfield  {journal} {\bibinfo  {journal} {Physical review letters}\ }\textbf {\bibinfo {volume} {113}},\ \bibinfo {pages} {026803} (\bibinfo {year} {2014})}\BibitemShut {NoStop}%
\bibitem [{\citenamefont {Chernikov}\ \emph {et~al.}(2014)\citenamefont {Chernikov}, \citenamefont {Berkelbach}, \citenamefont {Hill}, \citenamefont {Rigosi}, \citenamefont {Li}, \citenamefont {Aslan}, \citenamefont {Reichman}, \citenamefont {Hybertsen},\ and\ \citenamefont {Heinz}}]{4}%
  \BibitemOpen
  \bibfield  {author} {\bibinfo {author} {\bibfnamefont {A.}~\bibnamefont {Chernikov}}, \bibinfo {author} {\bibfnamefont {T.~C.}\ \bibnamefont {Berkelbach}}, \bibinfo {author} {\bibfnamefont {H.~M.}\ \bibnamefont {Hill}}, \bibinfo {author} {\bibfnamefont {A.}~\bibnamefont {Rigosi}}, \bibinfo {author} {\bibfnamefont {Y.}~\bibnamefont {Li}}, \bibinfo {author} {\bibfnamefont {B.}~\bibnamefont {Aslan}}, \bibinfo {author} {\bibfnamefont {D.~R.}\ \bibnamefont {Reichman}}, \bibinfo {author} {\bibfnamefont {M.~S.}\ \bibnamefont {Hybertsen}}, \ and\ \bibinfo {author} {\bibfnamefont {T.~F.}\ \bibnamefont {Heinz}},\ }\href@noop {} {\bibfield  {journal} {\bibinfo  {journal} {Physical review letters}\ }\textbf {\bibinfo {volume} {113}},\ \bibinfo {pages} {076802} (\bibinfo {year} {2014})}\BibitemShut {NoStop}%
\bibitem [{\citenamefont {Mak}\ \emph {et~al.}(2013)\citenamefont {Mak}, \citenamefont {He}, \citenamefont {Lee}, \citenamefont {Lee}, \citenamefont {Hone}, \citenamefont {Heinz},\ and\ \citenamefont {Shan}}]{5}%
  \BibitemOpen
  \bibfield  {author} {\bibinfo {author} {\bibfnamefont {K.~F.}\ \bibnamefont {Mak}}, \bibinfo {author} {\bibfnamefont {K.}~\bibnamefont {He}}, \bibinfo {author} {\bibfnamefont {C.}~\bibnamefont {Lee}}, \bibinfo {author} {\bibfnamefont {G.~H.}\ \bibnamefont {Lee}}, \bibinfo {author} {\bibfnamefont {J.}~\bibnamefont {Hone}}, \bibinfo {author} {\bibfnamefont {T.~F.}\ \bibnamefont {Heinz}}, \ and\ \bibinfo {author} {\bibfnamefont {J.}~\bibnamefont {Shan}},\ }\href {\doibase 10.1038/nmat3505} {\bibfield  {journal} {\bibinfo  {journal} {Nature Materials}\ }\textbf {\bibinfo {volume} {12}},\ \bibinfo {pages} {207} (\bibinfo {year} {2013})}\BibitemShut {NoStop}%
\bibitem [{\citenamefont {Rosati}\ \emph {et~al.}(2021)\citenamefont {Rosati}, \citenamefont {Schmidt}, \citenamefont {Brem}, \citenamefont {Perea-Causín}, \citenamefont {Niehues}, \citenamefont {Kern}, \citenamefont {Preuß}, \citenamefont {Schneider}, \citenamefont {Michaelis~de Vasconcellos}, \citenamefont {Bratschitsch},\ and\ \citenamefont {Malic}}]{6}%
  \BibitemOpen
  \bibfield  {author} {\bibinfo {author} {\bibfnamefont {R.}~\bibnamefont {Rosati}}, \bibinfo {author} {\bibfnamefont {R.}~\bibnamefont {Schmidt}}, \bibinfo {author} {\bibfnamefont {S.}~\bibnamefont {Brem}}, \bibinfo {author} {\bibfnamefont {R.}~\bibnamefont {Perea-Causín}}, \bibinfo {author} {\bibfnamefont {I.}~\bibnamefont {Niehues}}, \bibinfo {author} {\bibfnamefont {J.}~\bibnamefont {Kern}}, \bibinfo {author} {\bibfnamefont {J.~A.}\ \bibnamefont {Preuß}}, \bibinfo {author} {\bibfnamefont {R.}~\bibnamefont {Schneider}}, \bibinfo {author} {\bibfnamefont {S.}~\bibnamefont {Michaelis~de Vasconcellos}}, \bibinfo {author} {\bibfnamefont {R.}~\bibnamefont {Bratschitsch}}, \ and\ \bibinfo {author} {\bibfnamefont {E.}~\bibnamefont {Malic}},\ }\href {\doibase 10.1038/s41467-021-27425-y} {\bibfield  {journal} {\bibinfo  {journal} {Nature Communications}\ }\textbf {\bibinfo {volume} {12}} (\bibinfo {year} {2021}),\ 10.1038/s41467-021-27425-y}\BibitemShut {NoStop}%
\bibitem [{\citenamefont {Selig}\ \emph {et~al.}()\citenamefont {Selig}, \citenamefont {Berghäuser}, \citenamefont {Raja}, \citenamefont {Nagler}, \citenamefont {Schüller}, \citenamefont {Heinz}, \citenamefont {Korn}, \citenamefont {Chernikov}, \citenamefont {Malic},\ and\ \citenamefont {Knorr}}]{7}%
  \BibitemOpen
  \bibfield  {author} {\bibinfo {author} {\bibfnamefont {M.}~\bibnamefont {Selig}}, \bibinfo {author} {\bibfnamefont {G.}~\bibnamefont {Berghäuser}}, \bibinfo {author} {\bibfnamefont {A.}~\bibnamefont {Raja}}, \bibinfo {author} {\bibfnamefont {P.}~\bibnamefont {Nagler}}, \bibinfo {author} {\bibfnamefont {C.}~\bibnamefont {Schüller}}, \bibinfo {author} {\bibfnamefont {T.~F.}\ \bibnamefont {Heinz}}, \bibinfo {author} {\bibfnamefont {T.}~\bibnamefont {Korn}}, \bibinfo {author} {\bibfnamefont {A.}~\bibnamefont {Chernikov}}, \bibinfo {author} {\bibfnamefont {E.}~\bibnamefont {Malic}}, \ and\ \bibinfo {author} {\bibfnamefont {A.}~\bibnamefont {Knorr}},\ }\href {\doibase 10.1038/ncomms13279} {\bibfield  {journal} {\bibinfo  {journal} {Nature Communications}\ }\textbf {\bibinfo {volume} {7}},\ 10.1038/ncomms13279}\BibitemShut {NoStop}%
\bibitem [{\citenamefont {Selig}\ \emph {et~al.}(2018)\citenamefont {Selig}, \citenamefont {Bergh{\"a}user}, \citenamefont {Richter}, \citenamefont {Bratschitsch}, \citenamefont {Knorr},\ and\ \citenamefont {Malic}}]{8}%
  \BibitemOpen
  \bibfield  {author} {\bibinfo {author} {\bibfnamefont {M.}~\bibnamefont {Selig}}, \bibinfo {author} {\bibfnamefont {G.}~\bibnamefont {Bergh{\"a}user}}, \bibinfo {author} {\bibfnamefont {M.}~\bibnamefont {Richter}}, \bibinfo {author} {\bibfnamefont {R.}~\bibnamefont {Bratschitsch}}, \bibinfo {author} {\bibfnamefont {A.}~\bibnamefont {Knorr}}, \ and\ \bibinfo {author} {\bibfnamefont {E.}~\bibnamefont {Malic}},\ }\href@noop {} {\bibfield  {journal} {\bibinfo  {journal} {2D Materials}\ }\textbf {\bibinfo {volume} {5}},\ \bibinfo {pages} {035017} (\bibinfo {year} {2018})}\BibitemShut {NoStop}%
\bibitem [{\citenamefont {Lindlau}\ \emph {et~al.}(2018)\citenamefont {Lindlau}, \citenamefont {Selig}, \citenamefont {Neumann}, \citenamefont {Colombier}, \citenamefont {Förste}, \citenamefont {Funk}, \citenamefont {Förg}, \citenamefont {Kim}, \citenamefont {Berghäuser}, \citenamefont {Taniguchi}, \citenamefont {Watanabe}, \citenamefont {Wang}, \citenamefont {Malic},\ and\ \citenamefont {Högele}}]{9}%
  \BibitemOpen
  \bibfield  {author} {\bibinfo {author} {\bibfnamefont {J.}~\bibnamefont {Lindlau}}, \bibinfo {author} {\bibfnamefont {M.}~\bibnamefont {Selig}}, \bibinfo {author} {\bibfnamefont {A.}~\bibnamefont {Neumann}}, \bibinfo {author} {\bibfnamefont {L.}~\bibnamefont {Colombier}}, \bibinfo {author} {\bibfnamefont {J.}~\bibnamefont {Förste}}, \bibinfo {author} {\bibfnamefont {V.}~\bibnamefont {Funk}}, \bibinfo {author} {\bibfnamefont {M.}~\bibnamefont {Förg}}, \bibinfo {author} {\bibfnamefont {J.}~\bibnamefont {Kim}}, \bibinfo {author} {\bibfnamefont {G.}~\bibnamefont {Berghäuser}}, \bibinfo {author} {\bibfnamefont {T.}~\bibnamefont {Taniguchi}}, \bibinfo {author} {\bibfnamefont {K.}~\bibnamefont {Watanabe}}, \bibinfo {author} {\bibfnamefont {F.}~\bibnamefont {Wang}}, \bibinfo {author} {\bibfnamefont {E.}~\bibnamefont {Malic}}, \ and\ \bibinfo {author} {\bibfnamefont {A.}~\bibnamefont {Högele}},\ }\href {\doibase 10.1038/s41467-018-04877-3} {\bibfield  {journal} {\bibinfo  {journal} {Nature Communications}\
  }\textbf {\bibinfo {volume} {9}},\ \bibinfo {pages} {2586} (\bibinfo {year} {2018})}\BibitemShut {NoStop}%
\bibitem [{\citenamefont {Zipfel}\ \emph {et~al.}(2020)\citenamefont {Zipfel}, \citenamefont {Kulig}, \citenamefont {Perea-Caus{\'\i}n}, \citenamefont {Brem}, \citenamefont {Ziegler}, \citenamefont {Rosati}, \citenamefont {Taniguchi}, \citenamefont {Watanabe}, \citenamefont {Glazov}, \citenamefont {Malic} \emph {et~al.}}]{10}%
  \BibitemOpen
  \bibfield  {author} {\bibinfo {author} {\bibfnamefont {J.}~\bibnamefont {Zipfel}}, \bibinfo {author} {\bibfnamefont {M.}~\bibnamefont {Kulig}}, \bibinfo {author} {\bibfnamefont {R.}~\bibnamefont {Perea-Caus{\'\i}n}}, \bibinfo {author} {\bibfnamefont {S.}~\bibnamefont {Brem}}, \bibinfo {author} {\bibfnamefont {J.~D.}\ \bibnamefont {Ziegler}}, \bibinfo {author} {\bibfnamefont {R.}~\bibnamefont {Rosati}}, \bibinfo {author} {\bibfnamefont {T.}~\bibnamefont {Taniguchi}}, \bibinfo {author} {\bibfnamefont {K.}~\bibnamefont {Watanabe}}, \bibinfo {author} {\bibfnamefont {M.~M.}\ \bibnamefont {Glazov}}, \bibinfo {author} {\bibfnamefont {E.}~\bibnamefont {Malic}},  \emph {et~al.},\ }\href@noop {} {\bibfield  {journal} {\bibinfo  {journal} {Physical Review B}\ }\textbf {\bibinfo {volume} {101}},\ \bibinfo {pages} {115430} (\bibinfo {year} {2020})}\BibitemShut {NoStop}%
\bibitem [{\citenamefont {Chand}\ \emph {et~al.}(2023)\citenamefont {Chand}, \citenamefont {Woods}, \citenamefont {Quan}, \citenamefont {Mejia}, \citenamefont {Taniguchi}, \citenamefont {Watanabe}, \citenamefont {Alù},\ and\ \citenamefont {Grosso}}]{11}%
  \BibitemOpen
  \bibfield  {author} {\bibinfo {author} {\bibfnamefont {S.~B.}\ \bibnamefont {Chand}}, \bibinfo {author} {\bibfnamefont {J.~M.}\ \bibnamefont {Woods}}, \bibinfo {author} {\bibfnamefont {J.}~\bibnamefont {Quan}}, \bibinfo {author} {\bibfnamefont {E.}~\bibnamefont {Mejia}}, \bibinfo {author} {\bibfnamefont {T.}~\bibnamefont {Taniguchi}}, \bibinfo {author} {\bibfnamefont {K.}~\bibnamefont {Watanabe}}, \bibinfo {author} {\bibfnamefont {A.}~\bibnamefont {Alù}}, \ and\ \bibinfo {author} {\bibfnamefont {G.}~\bibnamefont {Grosso}},\ }\href {\doibase 10.1038/s41467-023-39339-y} {\bibfield  {journal} {\bibinfo  {journal} {Nature Communications}\ }\textbf {\bibinfo {volume} {14}},\ \bibinfo {pages} {3712} (\bibinfo {year} {2023})}\BibitemShut {NoStop}%
\bibitem [{\citenamefont {Rivera}\ \emph {et~al.}(2018)\citenamefont {Rivera}, \citenamefont {Yu}, \citenamefont {Seyler}, \citenamefont {Wilson}, \citenamefont {Yao},\ and\ \citenamefont {Xu}}]{12}%
  \BibitemOpen
  \bibfield  {author} {\bibinfo {author} {\bibfnamefont {P.}~\bibnamefont {Rivera}}, \bibinfo {author} {\bibfnamefont {H.}~\bibnamefont {Yu}}, \bibinfo {author} {\bibfnamefont {K.~L.}\ \bibnamefont {Seyler}}, \bibinfo {author} {\bibfnamefont {N.~P.}\ \bibnamefont {Wilson}}, \bibinfo {author} {\bibfnamefont {W.}~\bibnamefont {Yao}}, \ and\ \bibinfo {author} {\bibfnamefont {X.}~\bibnamefont {Xu}},\ }\href {\doibase 10.1038/s41565-018-0193-0} {\bibfield  {journal} {\bibinfo  {journal} {Nature Nanotechnology}\ }\textbf {\bibinfo {volume} {13}},\ \bibinfo {pages} {1004} (\bibinfo {year} {2018})}\BibitemShut {NoStop}%
\bibitem [{\citenamefont {Yagodkin}\ \emph {et~al.}(2023)\citenamefont {Yagodkin}, \citenamefont {Kumar}, \citenamefont {Ankerhold}, \citenamefont {Richter}, \citenamefont {Watanabe}, \citenamefont {Taniguchi}, \citenamefont {Gahl},\ and\ \citenamefont {Bolotin}}]{13}%
  \BibitemOpen
  \bibfield  {author} {\bibinfo {author} {\bibfnamefont {D.}~\bibnamefont {Yagodkin}}, \bibinfo {author} {\bibfnamefont {A.}~\bibnamefont {Kumar}}, \bibinfo {author} {\bibfnamefont {E.}~\bibnamefont {Ankerhold}}, \bibinfo {author} {\bibfnamefont {J.}~\bibnamefont {Richter}}, \bibinfo {author} {\bibfnamefont {K.}~\bibnamefont {Watanabe}}, \bibinfo {author} {\bibfnamefont {T.}~\bibnamefont {Taniguchi}}, \bibinfo {author} {\bibfnamefont {C.}~\bibnamefont {Gahl}}, \ and\ \bibinfo {author} {\bibfnamefont {K.~I.}\ \bibnamefont {Bolotin}},\ }\href {\doibase 10.1021/acs.nanolett.3c01708} {\bibfield  {journal} {\bibinfo  {journal} {Nano Letters}\ }\textbf {\bibinfo {volume} {23}} (\bibinfo {year} {2023}),\ 10.1021/acs.nanolett.3c01708}\BibitemShut {NoStop}%
\bibitem [{\citenamefont {Linhart}\ \emph {et~al.}(2019)\citenamefont {Linhart}, \citenamefont {Paur}, \citenamefont {Smejkal}, \citenamefont {Burgdörfer}, \citenamefont {Mueller},\ and\ \citenamefont {Libisch}}]{14}%
  \BibitemOpen
  \bibfield  {author} {\bibinfo {author} {\bibfnamefont {L.}~\bibnamefont {Linhart}}, \bibinfo {author} {\bibfnamefont {M.}~\bibnamefont {Paur}}, \bibinfo {author} {\bibfnamefont {V.}~\bibnamefont {Smejkal}}, \bibinfo {author} {\bibfnamefont {J.}~\bibnamefont {Burgdörfer}}, \bibinfo {author} {\bibfnamefont {T.}~\bibnamefont {Mueller}}, \ and\ \bibinfo {author} {\bibfnamefont {F.}~\bibnamefont {Libisch}},\ }\href {\doibase 10.1103/PhysRevLett.123.146401} {\bibfield  {journal} {\bibinfo  {journal} {Physical Review Letters}\ }\textbf {\bibinfo {volume} {123}} (\bibinfo {year} {2019}),\ 10.1103/PhysRevLett.123.146401}\BibitemShut {NoStop}%
\bibitem [{\citenamefont {Yagodkin}\ \emph {et~al.}(2022)\citenamefont {Yagodkin}, \citenamefont {Greben}, \citenamefont {Eljarrat}, \citenamefont {Kovalchuk}, \citenamefont {Ghorbani‐Asl}, \citenamefont {Jain}, \citenamefont {Kretschmer}, \citenamefont {Severin}, \citenamefont {Rabe}, \citenamefont {Krasheninnikov}, \citenamefont {Koch},\ and\ \citenamefont {Bolotin}}]{15}%
  \BibitemOpen
  \bibfield  {author} {\bibinfo {author} {\bibfnamefont {D.}~\bibnamefont {Yagodkin}}, \bibinfo {author} {\bibfnamefont {K.}~\bibnamefont {Greben}}, \bibinfo {author} {\bibfnamefont {A.}~\bibnamefont {Eljarrat}}, \bibinfo {author} {\bibfnamefont {S.}~\bibnamefont {Kovalchuk}}, \bibinfo {author} {\bibfnamefont {M.}~\bibnamefont {Ghorbani‐Asl}}, \bibinfo {author} {\bibfnamefont {M.}~\bibnamefont {Jain}}, \bibinfo {author} {\bibfnamefont {S.}~\bibnamefont {Kretschmer}}, \bibinfo {author} {\bibfnamefont {N.}~\bibnamefont {Severin}}, \bibinfo {author} {\bibfnamefont {J.~P.}\ \bibnamefont {Rabe}}, \bibinfo {author} {\bibfnamefont {A.~V.}\ \bibnamefont {Krasheninnikov}}, \bibinfo {author} {\bibfnamefont {C.~T.}\ \bibnamefont {Koch}}, \ and\ \bibinfo {author} {\bibfnamefont {K.~I.}\ \bibnamefont {Bolotin}},\ }\href {\doibase 10.1002/adfm.202203060} {\bibfield  {journal} {\bibinfo  {journal} {Advanced Functional Materials}\ }\textbf {\bibinfo {volume} {32}} (\bibinfo {year} {2022}),\
  10.1002/adfm.202203060}\BibitemShut {NoStop}%
\bibitem [{\citenamefont {Greben}\ \emph {et~al.}(2020)\citenamefont {Greben}, \citenamefont {Arora}, \citenamefont {Harats},\ and\ \citenamefont {Bolotin}}]{16}%
  \BibitemOpen
  \bibfield  {author} {\bibinfo {author} {\bibfnamefont {K.}~\bibnamefont {Greben}}, \bibinfo {author} {\bibfnamefont {S.}~\bibnamefont {Arora}}, \bibinfo {author} {\bibfnamefont {M.~G.}\ \bibnamefont {Harats}}, \ and\ \bibinfo {author} {\bibfnamefont {K.~I.}\ \bibnamefont {Bolotin}},\ }\href {\doibase 10.1021/acs.nanolett.9b05323} {\bibfield  {journal} {\bibinfo  {journal} {Nano Letters}\ }\textbf {\bibinfo {volume} {20}},\ \bibinfo {pages} {2544} (\bibinfo {year} {2020})}\BibitemShut {NoStop}%
\bibitem [{\citenamefont {Deilmann}\ and\ \citenamefont {Thygesen}(2019)}]{17}%
  \BibitemOpen
  \bibfield  {author} {\bibinfo {author} {\bibfnamefont {T.}~\bibnamefont {Deilmann}}\ and\ \bibinfo {author} {\bibfnamefont {K.~S.}\ \bibnamefont {Thygesen}},\ }\href@noop {} {\bibfield  {journal} {\bibinfo  {journal} {2D Materials}\ }\textbf {\bibinfo {volume} {6}},\ \bibinfo {pages} {035003} (\bibinfo {year} {2019})}\BibitemShut {NoStop}%
\bibitem [{\citenamefont {Madéo}\ \emph {et~al.}(2020)\citenamefont {Madéo}, \citenamefont {Man}, \citenamefont {Sahoo}, \citenamefont {Campbell}, \citenamefont {Pareek}, \citenamefont {Wong}, \citenamefont {Al-Mahboob}, \citenamefont {Chan}, \citenamefont {Karmakar}, \citenamefont {Mariserla}, \citenamefont {Li}, \citenamefont {Heinz}, \citenamefont {Cao},\ and\ \citenamefont {Dani}}]{18}%
  \BibitemOpen
  \bibfield  {author} {\bibinfo {author} {\bibfnamefont {J.}~\bibnamefont {Madéo}}, \bibinfo {author} {\bibfnamefont {M.~K.}\ \bibnamefont {Man}}, \bibinfo {author} {\bibfnamefont {C.}~\bibnamefont {Sahoo}}, \bibinfo {author} {\bibfnamefont {M.}~\bibnamefont {Campbell}}, \bibinfo {author} {\bibfnamefont {V.}~\bibnamefont {Pareek}}, \bibinfo {author} {\bibfnamefont {E.~L.}\ \bibnamefont {Wong}}, \bibinfo {author} {\bibfnamefont {A.}~\bibnamefont {Al-Mahboob}}, \bibinfo {author} {\bibfnamefont {N.~S.}\ \bibnamefont {Chan}}, \bibinfo {author} {\bibfnamefont {A.}~\bibnamefont {Karmakar}}, \bibinfo {author} {\bibfnamefont {B.~M.~K.}\ \bibnamefont {Mariserla}}, \bibinfo {author} {\bibfnamefont {X.}~\bibnamefont {Li}}, \bibinfo {author} {\bibfnamefont {T.~F.}\ \bibnamefont {Heinz}}, \bibinfo {author} {\bibfnamefont {T.}~\bibnamefont {Cao}}, \ and\ \bibinfo {author} {\bibfnamefont {K.~M.}\ \bibnamefont {Dani}},\ }\href {\doibase 10.1126/science.aba1029} {\bibfield  {journal} {\bibinfo  {journal} {Science}\ }\textbf
  {\bibinfo {volume} {370}} (\bibinfo {year} {2020}),\ 10.1126/science.aba1029}\BibitemShut {NoStop}%
\bibitem [{\citenamefont {Dong}\ \emph {et~al.}(2021)\citenamefont {Dong}, \citenamefont {Puppin}, \citenamefont {Pincelli}, \citenamefont {Beaulieu}, \citenamefont {Christiansen}, \citenamefont {Hübener}, \citenamefont {Nicholson}, \citenamefont {Xian}, \citenamefont {Dendzik}, \citenamefont {Deng}, \citenamefont {Windsor}, \citenamefont {Selig}, \citenamefont {Malic}, \citenamefont {Rubio}, \citenamefont {Knorr}, \citenamefont {Wolf}, \citenamefont {Rettig},\ and\ \citenamefont {Ernstorfer}}]{19}%
  \BibitemOpen
  \bibfield  {author} {\bibinfo {author} {\bibfnamefont {S.}~\bibnamefont {Dong}}, \bibinfo {author} {\bibfnamefont {M.}~\bibnamefont {Puppin}}, \bibinfo {author} {\bibfnamefont {T.}~\bibnamefont {Pincelli}}, \bibinfo {author} {\bibfnamefont {S.}~\bibnamefont {Beaulieu}}, \bibinfo {author} {\bibfnamefont {D.}~\bibnamefont {Christiansen}}, \bibinfo {author} {\bibfnamefont {H.}~\bibnamefont {Hübener}}, \bibinfo {author} {\bibfnamefont {C.~W.}\ \bibnamefont {Nicholson}}, \bibinfo {author} {\bibfnamefont {R.~P.}\ \bibnamefont {Xian}}, \bibinfo {author} {\bibfnamefont {M.}~\bibnamefont {Dendzik}}, \bibinfo {author} {\bibfnamefont {Y.}~\bibnamefont {Deng}}, \bibinfo {author} {\bibfnamefont {Y.~W.}\ \bibnamefont {Windsor}}, \bibinfo {author} {\bibfnamefont {M.}~\bibnamefont {Selig}}, \bibinfo {author} {\bibfnamefont {E.}~\bibnamefont {Malic}}, \bibinfo {author} {\bibfnamefont {A.}~\bibnamefont {Rubio}}, \bibinfo {author} {\bibfnamefont {A.}~\bibnamefont {Knorr}}, \bibinfo {author} {\bibfnamefont {M.}~\bibnamefont
  {Wolf}}, \bibinfo {author} {\bibfnamefont {L.}~\bibnamefont {Rettig}}, \ and\ \bibinfo {author} {\bibfnamefont {R.}~\bibnamefont {Ernstorfer}},\ }\href {\doibase 10.1002/ntls.10010} {\bibfield  {journal} {\bibinfo  {journal} {Natural Sciences}\ }\textbf {\bibinfo {volume} {1}},\ \bibinfo {pages} {e10010} (\bibinfo {year} {2021})}\BibitemShut {NoStop}%
\bibitem [{\citenamefont {Hong}\ \emph {et~al.}(2020)\citenamefont {Hong}, \citenamefont {Senga}, \citenamefont {Pichler},\ and\ \citenamefont {Suenaga}}]{20}%
  \BibitemOpen
  \bibfield  {author} {\bibinfo {author} {\bibfnamefont {J.}~\bibnamefont {Hong}}, \bibinfo {author} {\bibfnamefont {R.}~\bibnamefont {Senga}}, \bibinfo {author} {\bibfnamefont {T.}~\bibnamefont {Pichler}}, \ and\ \bibinfo {author} {\bibfnamefont {K.}~\bibnamefont {Suenaga}},\ }\href {\doibase 10.1103/PhysRevLett.124.087401} {\bibfield  {journal} {\bibinfo  {journal} {Physical Review Letters}\ }\textbf {\bibinfo {volume} {124}} (\bibinfo {year} {2020}),\ 10.1103/PhysRevLett.124.087401}\BibitemShut {NoStop}%
\bibitem [{\citenamefont {Schmitt}\ \emph {et~al.}(2022)\citenamefont {Schmitt}, \citenamefont {Bange}, \citenamefont {Bennecke}, \citenamefont {AlMutairi}, \citenamefont {Meneghini}, \citenamefont {Watanabe}, \citenamefont {Taniguchi}, \citenamefont {Steil}, \citenamefont {Luke}, \citenamefont {Weitz}, \citenamefont {Steil}, \citenamefont {Jansen}, \citenamefont {Brem}, \citenamefont {Malic}, \citenamefont {Hofmann}, \citenamefont {Reutzel},\ and\ \citenamefont {Mathias}}]{21}%
  \BibitemOpen
  \bibfield  {author} {\bibinfo {author} {\bibfnamefont {D.}~\bibnamefont {Schmitt}}, \bibinfo {author} {\bibfnamefont {J.~P.}\ \bibnamefont {Bange}}, \bibinfo {author} {\bibfnamefont {W.}~\bibnamefont {Bennecke}}, \bibinfo {author} {\bibfnamefont {A.}~\bibnamefont {AlMutairi}}, \bibinfo {author} {\bibfnamefont {G.}~\bibnamefont {Meneghini}}, \bibinfo {author} {\bibfnamefont {K.}~\bibnamefont {Watanabe}}, \bibinfo {author} {\bibfnamefont {T.}~\bibnamefont {Taniguchi}}, \bibinfo {author} {\bibfnamefont {D.}~\bibnamefont {Steil}}, \bibinfo {author} {\bibfnamefont {D.~R.}\ \bibnamefont {Luke}}, \bibinfo {author} {\bibfnamefont {R.~T.}\ \bibnamefont {Weitz}}, \bibinfo {author} {\bibfnamefont {S.}~\bibnamefont {Steil}}, \bibinfo {author} {\bibfnamefont {G.~S.~M.}\ \bibnamefont {Jansen}}, \bibinfo {author} {\bibfnamefont {S.}~\bibnamefont {Brem}}, \bibinfo {author} {\bibfnamefont {E.}~\bibnamefont {Malic}}, \bibinfo {author} {\bibfnamefont {S.}~\bibnamefont {Hofmann}}, \bibinfo {author} {\bibfnamefont
  {M.}~\bibnamefont {Reutzel}}, \ and\ \bibinfo {author} {\bibfnamefont {S.}~\bibnamefont {Mathias}},\ }\href {\doibase 10.1038/s41586-022-04977-7} {\bibfield  {journal} {\bibinfo  {journal} {Nature}\ }\textbf {\bibinfo {volume} {608}} (\bibinfo {year} {2022}),\ 10.1038/s41586-022-04977-7}\BibitemShut {NoStop}%
\bibitem [{\citenamefont {Wallauer}\ \emph {et~al.}(2021)\citenamefont {Wallauer}, \citenamefont {Perea-Causin}, \citenamefont {Münster}, \citenamefont {Zajusch}, \citenamefont {Brem}, \citenamefont {Güdde}, \citenamefont {Tanimura}, \citenamefont {Lin}, \citenamefont {Huber}, \citenamefont {Malic},\ and\ \citenamefont {Höfer}}]{22}%
  \BibitemOpen
  \bibfield  {author} {\bibinfo {author} {\bibfnamefont {R.}~\bibnamefont {Wallauer}}, \bibinfo {author} {\bibfnamefont {R.}~\bibnamefont {Perea-Causin}}, \bibinfo {author} {\bibfnamefont {L.}~\bibnamefont {Münster}}, \bibinfo {author} {\bibfnamefont {S.}~\bibnamefont {Zajusch}}, \bibinfo {author} {\bibfnamefont {S.}~\bibnamefont {Brem}}, \bibinfo {author} {\bibfnamefont {J.}~\bibnamefont {Güdde}}, \bibinfo {author} {\bibfnamefont {K.}~\bibnamefont {Tanimura}}, \bibinfo {author} {\bibfnamefont {K.-Q.}\ \bibnamefont {Lin}}, \bibinfo {author} {\bibfnamefont {R.}~\bibnamefont {Huber}}, \bibinfo {author} {\bibfnamefont {E.}~\bibnamefont {Malic}}, \ and\ \bibinfo {author} {\bibfnamefont {U.}~\bibnamefont {Höfer}},\ }\href {\doibase 10.1021/acs.nanolett.1c01839} {\bibfield  {journal} {\bibinfo  {journal} {Nano Letters}\ }\textbf {\bibinfo {volume} {21}},\ \bibinfo {pages} {5867} (\bibinfo {year} {2021})}\BibitemShut {NoStop}%
\bibitem [{\citenamefont {Feierabend}\ \emph {et~al.}(2017)\citenamefont {Feierabend}, \citenamefont {Morlet}, \citenamefont {Berghäuser},\ and\ \citenamefont {Malic}}]{23}%
  \BibitemOpen
  \bibfield  {author} {\bibinfo {author} {\bibfnamefont {M.}~\bibnamefont {Feierabend}}, \bibinfo {author} {\bibfnamefont {A.}~\bibnamefont {Morlet}}, \bibinfo {author} {\bibfnamefont {G.}~\bibnamefont {Berghäuser}}, \ and\ \bibinfo {author} {\bibfnamefont {E.}~\bibnamefont {Malic}},\ }\href {\doibase 10.1103/PhysRevB.96.045425} {\bibfield  {journal} {\bibinfo  {journal} {Physical Review B}\ }\textbf {\bibinfo {volume} {96}} (\bibinfo {year} {2017}),\ 10.1103/PhysRevB.96.045425}\BibitemShut {NoStop}%
\bibitem [{\citenamefont {Zollner}\ \emph {et~al.}(2019)\citenamefont {Zollner}, \citenamefont {Junior},\ and\ \citenamefont {Fabian}}]{24}%
  \BibitemOpen
  \bibfield  {author} {\bibinfo {author} {\bibfnamefont {K.}~\bibnamefont {Zollner}}, \bibinfo {author} {\bibfnamefont {P.~E.~F.}\ \bibnamefont {Junior}}, \ and\ \bibinfo {author} {\bibfnamefont {J.}~\bibnamefont {Fabian}},\ }\href {\doibase 10.1103/PhysRevB.100.195126} {\bibfield  {journal} {\bibinfo  {journal} {Physical Review B}\ }\textbf {\bibinfo {volume} {100}} (\bibinfo {year} {2019}),\ 10.1103/PhysRevB.100.195126}\BibitemShut {NoStop}%
\bibitem [{\citenamefont {Defo}\ \emph {et~al.}(2016)\citenamefont {Defo}, \citenamefont {Fang}, \citenamefont {Shirodkar}, \citenamefont {Tritsaris}, \citenamefont {Dimoulas},\ and\ \citenamefont {Kaxiras}}]{25}%
  \BibitemOpen
  \bibfield  {author} {\bibinfo {author} {\bibfnamefont {R.~K.}\ \bibnamefont {Defo}}, \bibinfo {author} {\bibfnamefont {S.}~\bibnamefont {Fang}}, \bibinfo {author} {\bibfnamefont {S.~N.}\ \bibnamefont {Shirodkar}}, \bibinfo {author} {\bibfnamefont {G.~A.}\ \bibnamefont {Tritsaris}}, \bibinfo {author} {\bibfnamefont {A.}~\bibnamefont {Dimoulas}}, \ and\ \bibinfo {author} {\bibfnamefont {E.}~\bibnamefont {Kaxiras}},\ }\href {\doibase 10.1103/PhysRevB.94.155310} {\bibfield  {journal} {\bibinfo  {journal} {Physical Review B}\ }\textbf {\bibinfo {volume} {94}} (\bibinfo {year} {2016}),\ 10.1103/PhysRevB.94.155310}\BibitemShut {NoStop}%
\bibitem [{\citenamefont {Khatibi}\ \emph {et~al.}(2019)\citenamefont {Khatibi}, \citenamefont {Feierabend}, \citenamefont {Selig}, \citenamefont {Brem}, \citenamefont {Linderälv}, \citenamefont {Erhart},\ and\ \citenamefont {Malic}}]{26}%
  \BibitemOpen
  \bibfield  {author} {\bibinfo {author} {\bibfnamefont {Z.}~\bibnamefont {Khatibi}}, \bibinfo {author} {\bibfnamefont {M.}~\bibnamefont {Feierabend}}, \bibinfo {author} {\bibfnamefont {M.}~\bibnamefont {Selig}}, \bibinfo {author} {\bibfnamefont {S.}~\bibnamefont {Brem}}, \bibinfo {author} {\bibfnamefont {C.}~\bibnamefont {Linderälv}}, \bibinfo {author} {\bibfnamefont {P.}~\bibnamefont {Erhart}}, \ and\ \bibinfo {author} {\bibfnamefont {E.}~\bibnamefont {Malic}},\ }\href {\doibase 10.1088/2053-1583/aae953} {\bibfield  {journal} {\bibinfo  {journal} {2D Materials}\ }\textbf {\bibinfo {volume} {6}} (\bibinfo {year} {2019}),\ 10.1088/2053-1583/aae953}\BibitemShut {NoStop}%
\bibitem [{\citenamefont {Hernández~López}\ \emph {et~al.}(2022)\citenamefont {Hernández~López}, \citenamefont {Heeg}, \citenamefont {Schattauer}, \citenamefont {Kovalchuk}, \citenamefont {Kumar}, \citenamefont {Bock}, \citenamefont {Kirchhof}, \citenamefont {Höfer}, \citenamefont {Greben}, \citenamefont {Yagodkin}, \citenamefont {Linhart}, \citenamefont {Libisch},\ and\ \citenamefont {Bolotin}}]{27}%
  \BibitemOpen
  \bibfield  {author} {\bibinfo {author} {\bibfnamefont {P.}~\bibnamefont {Hernández~López}}, \bibinfo {author} {\bibfnamefont {S.}~\bibnamefont {Heeg}}, \bibinfo {author} {\bibfnamefont {C.}~\bibnamefont {Schattauer}}, \bibinfo {author} {\bibfnamefont {S.}~\bibnamefont {Kovalchuk}}, \bibinfo {author} {\bibfnamefont {A.}~\bibnamefont {Kumar}}, \bibinfo {author} {\bibfnamefont {D.~J.}\ \bibnamefont {Bock}}, \bibinfo {author} {\bibfnamefont {J.~N.}\ \bibnamefont {Kirchhof}}, \bibinfo {author} {\bibfnamefont {B.}~\bibnamefont {Höfer}}, \bibinfo {author} {\bibfnamefont {K.}~\bibnamefont {Greben}}, \bibinfo {author} {\bibfnamefont {D.}~\bibnamefont {Yagodkin}}, \bibinfo {author} {\bibfnamefont {L.}~\bibnamefont {Linhart}}, \bibinfo {author} {\bibfnamefont {F.}~\bibnamefont {Libisch}}, \ and\ \bibinfo {author} {\bibfnamefont {K.~I.}\ \bibnamefont {Bolotin}},\ }\href {\doibase 10.1038/s41467-022-35352-9} {\bibfield  {journal} {\bibinfo  {journal} {Nature Communications}\ }\textbf {\bibinfo {volume} {13}},\
  \bibinfo {pages} {7691} (\bibinfo {year} {2022})}\BibitemShut {NoStop}%
\bibitem [{\citenamefont {Niehues}\ \emph {et~al.}(2018)\citenamefont {Niehues}, \citenamefont {Schmidt}, \citenamefont {Drüppel}, \citenamefont {Marauhn}, \citenamefont {Christiansen}, \citenamefont {Selig}, \citenamefont {Berghäuser}, \citenamefont {Wigger}, \citenamefont {Schneider}, \citenamefont {Braasch}, \citenamefont {Koch}, \citenamefont {Castellanos-Gomez}, \citenamefont {Kuhn}, \citenamefont {Knorr}, \citenamefont {Malic}, \citenamefont {Rohlfing}, \citenamefont {Michaelis~de Vasconcellos},\ and\ \citenamefont {Bratschitsch}}]{28}%
  \BibitemOpen
  \bibfield  {author} {\bibinfo {author} {\bibfnamefont {I.}~\bibnamefont {Niehues}}, \bibinfo {author} {\bibfnamefont {R.}~\bibnamefont {Schmidt}}, \bibinfo {author} {\bibfnamefont {M.}~\bibnamefont {Drüppel}}, \bibinfo {author} {\bibfnamefont {P.}~\bibnamefont {Marauhn}}, \bibinfo {author} {\bibfnamefont {D.}~\bibnamefont {Christiansen}}, \bibinfo {author} {\bibfnamefont {M.}~\bibnamefont {Selig}}, \bibinfo {author} {\bibfnamefont {G.}~\bibnamefont {Berghäuser}}, \bibinfo {author} {\bibfnamefont {D.}~\bibnamefont {Wigger}}, \bibinfo {author} {\bibfnamefont {R.}~\bibnamefont {Schneider}}, \bibinfo {author} {\bibfnamefont {L.}~\bibnamefont {Braasch}}, \bibinfo {author} {\bibfnamefont {R.}~\bibnamefont {Koch}}, \bibinfo {author} {\bibfnamefont {A.}~\bibnamefont {Castellanos-Gomez}}, \bibinfo {author} {\bibfnamefont {T.}~\bibnamefont {Kuhn}}, \bibinfo {author} {\bibfnamefont {A.}~\bibnamefont {Knorr}}, \bibinfo {author} {\bibfnamefont {E.}~\bibnamefont {Malic}}, \bibinfo {author} {\bibfnamefont
  {M.}~\bibnamefont {Rohlfing}}, \bibinfo {author} {\bibfnamefont {S.}~\bibnamefont {Michaelis~de Vasconcellos}}, \ and\ \bibinfo {author} {\bibfnamefont {R.}~\bibnamefont {Bratschitsch}},\ }\href {\doibase 10.1021/acs.nanolett.7b04868} {\bibfield  {journal} {\bibinfo  {journal} {Nano Letters}\ }\textbf {\bibinfo {volume} {18}} (\bibinfo {year} {2018}),\ 10.1021/acs.nanolett.7b04868}\BibitemShut {NoStop}%
\bibitem [{\citenamefont {Aslan}\ \emph {et~al.}(2021)\citenamefont {Aslan}, \citenamefont {Yule}, \citenamefont {Yu}, \citenamefont {Lee}, \citenamefont {Heinz}, \citenamefont {Cao},\ and\ \citenamefont {Brongersma}}]{29}%
  \BibitemOpen
  \bibfield  {author} {\bibinfo {author} {\bibfnamefont {B.}~\bibnamefont {Aslan}}, \bibinfo {author} {\bibfnamefont {C.}~\bibnamefont {Yule}}, \bibinfo {author} {\bibfnamefont {Y.}~\bibnamefont {Yu}}, \bibinfo {author} {\bibfnamefont {Y.~J.}\ \bibnamefont {Lee}}, \bibinfo {author} {\bibfnamefont {T.~F.}\ \bibnamefont {Heinz}}, \bibinfo {author} {\bibfnamefont {L.}~\bibnamefont {Cao}}, \ and\ \bibinfo {author} {\bibfnamefont {M.~L.}\ \bibnamefont {Brongersma}},\ }\href {\doibase 10.1088/2053-1583/ac2d15} {\bibfield  {journal} {\bibinfo  {journal} {2D Materials}\ }\textbf {\bibinfo {volume} {9}} (\bibinfo {year} {2021}),\ 10.1088/2053-1583/ac2d15}\BibitemShut {NoStop}%
\bibitem [{\citenamefont {Harats}\ \emph {et~al.}(2020)\citenamefont {Harats}, \citenamefont {Kirchhof}, \citenamefont {Qiao}, \citenamefont {Greben},\ and\ \citenamefont {Bolotin}}]{30}%
  \BibitemOpen
  \bibfield  {author} {\bibinfo {author} {\bibfnamefont {M.~G.}\ \bibnamefont {Harats}}, \bibinfo {author} {\bibfnamefont {J.~N.}\ \bibnamefont {Kirchhof}}, \bibinfo {author} {\bibfnamefont {M.}~\bibnamefont {Qiao}}, \bibinfo {author} {\bibfnamefont {K.}~\bibnamefont {Greben}}, \ and\ \bibinfo {author} {\bibfnamefont {K.~I.}\ \bibnamefont {Bolotin}},\ }\href {\doibase 10.1038/s41566-019-0581-5} {\bibfield  {journal} {\bibinfo  {journal} {Nature Photonics}\ }\textbf {\bibinfo {volume} {14}},\ \bibinfo {pages} {324} (\bibinfo {year} {2020})}\BibitemShut {NoStop}%
\bibitem [{\citenamefont {Schmidt}\ \emph {et~al.}(2016)\citenamefont {Schmidt}, \citenamefont {Niehues}, \citenamefont {Schneider}, \citenamefont {Drüppel}, \citenamefont {Deilmann}, \citenamefont {Rohlfing}, \citenamefont {Vasconcellos}, \citenamefont {Castellanos-Gomez},\ and\ \citenamefont {Bratschitsch}}]{31}%
  \BibitemOpen
  \bibfield  {author} {\bibinfo {author} {\bibfnamefont {R.}~\bibnamefont {Schmidt}}, \bibinfo {author} {\bibfnamefont {I.}~\bibnamefont {Niehues}}, \bibinfo {author} {\bibfnamefont {R.}~\bibnamefont {Schneider}}, \bibinfo {author} {\bibfnamefont {M.}~\bibnamefont {Drüppel}}, \bibinfo {author} {\bibfnamefont {T.}~\bibnamefont {Deilmann}}, \bibinfo {author} {\bibfnamefont {M.}~\bibnamefont {Rohlfing}}, \bibinfo {author} {\bibfnamefont {S.~M.~d.}\ \bibnamefont {Vasconcellos}}, \bibinfo {author} {\bibfnamefont {A.}~\bibnamefont {Castellanos-Gomez}}, \ and\ \bibinfo {author} {\bibfnamefont {R.}~\bibnamefont {Bratschitsch}},\ }\href {\doibase 10.1088/2053-1583/3/2/021011} {\bibfield  {journal} {\bibinfo  {journal} {2D Materials}\ }\textbf {\bibinfo {volume} {3}} (\bibinfo {year} {2016}),\ 10.1088/2053-1583/3/2/021011}\BibitemShut {NoStop}%
\bibitem [{\citenamefont {Blundo}\ \emph {et~al.}(2022)\citenamefont {Blundo}, \citenamefont {Junior}, \citenamefont {Surrente}, \citenamefont {Pettinari}, \citenamefont {Prosnikov}, \citenamefont {Olkowska-Pucko}, \citenamefont {Zollner}, \citenamefont {Wo{\'z}niak}, \citenamefont {Chaves}, \citenamefont {Kazimierczuk} \emph {et~al.}}]{32}%
  \BibitemOpen
  \bibfield  {author} {\bibinfo {author} {\bibfnamefont {E.}~\bibnamefont {Blundo}}, \bibinfo {author} {\bibfnamefont {P.~E.~F.}\ \bibnamefont {Junior}}, \bibinfo {author} {\bibfnamefont {A.}~\bibnamefont {Surrente}}, \bibinfo {author} {\bibfnamefont {G.}~\bibnamefont {Pettinari}}, \bibinfo {author} {\bibfnamefont {M.~A.}\ \bibnamefont {Prosnikov}}, \bibinfo {author} {\bibfnamefont {K.}~\bibnamefont {Olkowska-Pucko}}, \bibinfo {author} {\bibfnamefont {K.}~\bibnamefont {Zollner}}, \bibinfo {author} {\bibfnamefont {T.}~\bibnamefont {Wo{\'z}niak}}, \bibinfo {author} {\bibfnamefont {A.}~\bibnamefont {Chaves}}, \bibinfo {author} {\bibfnamefont {T.}~\bibnamefont {Kazimierczuk}},  \emph {et~al.},\ }\href@noop {} {\bibfield  {journal} {\bibinfo  {journal} {Physical Review Letters}\ }\textbf {\bibinfo {volume} {129}},\ \bibinfo {pages} {067402} (\bibinfo {year} {2022})}\BibitemShut {NoStop}%
\bibitem [{\citenamefont {Blundo}\ \emph {et~al.}(2020)\citenamefont {Blundo}, \citenamefont {Felici}, \citenamefont {Yildirim}, \citenamefont {Pettinari}, \citenamefont {Tedeschi}, \citenamefont {Miriametro}, \citenamefont {Liu}, \citenamefont {Ma}, \citenamefont {Lu},\ and\ \citenamefont {Polimeni}}]{33}%
  \BibitemOpen
  \bibfield  {author} {\bibinfo {author} {\bibfnamefont {E.}~\bibnamefont {Blundo}}, \bibinfo {author} {\bibfnamefont {M.}~\bibnamefont {Felici}}, \bibinfo {author} {\bibfnamefont {T.}~\bibnamefont {Yildirim}}, \bibinfo {author} {\bibfnamefont {G.}~\bibnamefont {Pettinari}}, \bibinfo {author} {\bibfnamefont {D.}~\bibnamefont {Tedeschi}}, \bibinfo {author} {\bibfnamefont {A.}~\bibnamefont {Miriametro}}, \bibinfo {author} {\bibfnamefont {B.}~\bibnamefont {Liu}}, \bibinfo {author} {\bibfnamefont {W.}~\bibnamefont {Ma}}, \bibinfo {author} {\bibfnamefont {Y.}~\bibnamefont {Lu}}, \ and\ \bibinfo {author} {\bibfnamefont {A.}~\bibnamefont {Polimeni}},\ }\href@noop {} {\bibfield  {journal} {\bibinfo  {journal} {Physical Review Research}\ }\textbf {\bibinfo {volume} {2}},\ \bibinfo {pages} {012024} (\bibinfo {year} {2020})}\BibitemShut {NoStop}%
\bibitem [{\citenamefont {Kovalchuk}\ \emph {et~al.}(2020)\citenamefont {Kovalchuk}, \citenamefont {Harats}, \citenamefont {López-Polín}, \citenamefont {Kirchhof}, \citenamefont {Höflich},\ and\ \citenamefont {Bolotin}}]{34}%
  \BibitemOpen
  \bibfield  {author} {\bibinfo {author} {\bibfnamefont {S.}~\bibnamefont {Kovalchuk}}, \bibinfo {author} {\bibfnamefont {M.~G.}\ \bibnamefont {Harats}}, \bibinfo {author} {\bibfnamefont {G.}~\bibnamefont {López-Polín}}, \bibinfo {author} {\bibfnamefont {J.~N.}\ \bibnamefont {Kirchhof}}, \bibinfo {author} {\bibfnamefont {K.}~\bibnamefont {Höflich}}, \ and\ \bibinfo {author} {\bibfnamefont {K.~I.}\ \bibnamefont {Bolotin}},\ }\href {\doibase 10.1088/2053-1583/ab8caa} {\bibfield  {journal} {\bibinfo  {journal} {2D Materials}\ }\textbf {\bibinfo {volume} {7}} (\bibinfo {year} {2020}),\ 10.1088/2053-1583/ab8caa}\BibitemShut {NoStop}%
\bibitem [{\citenamefont {Rivera}\ \emph {et~al.}(2021)\citenamefont {Rivera}, \citenamefont {He}, \citenamefont {Kim}, \citenamefont {Liu}, \citenamefont {Rubio-Verdú}, \citenamefont {Moon}, \citenamefont {Mennel}, \citenamefont {Rhodes}, \citenamefont {Yu}, \citenamefont {Taniguchi}, \citenamefont {Watanabe}, \citenamefont {Yan}, \citenamefont {Mandrus}, \citenamefont {Dery}, \citenamefont {Pasupathy}, \citenamefont {Englund}, \citenamefont {Hone}, \citenamefont {Yao},\ and\ \citenamefont {Xu}}]{35}%
  \BibitemOpen
  \bibfield  {author} {\bibinfo {author} {\bibfnamefont {P.}~\bibnamefont {Rivera}}, \bibinfo {author} {\bibfnamefont {M.}~\bibnamefont {He}}, \bibinfo {author} {\bibfnamefont {B.}~\bibnamefont {Kim}}, \bibinfo {author} {\bibfnamefont {S.}~\bibnamefont {Liu}}, \bibinfo {author} {\bibfnamefont {C.}~\bibnamefont {Rubio-Verdú}}, \bibinfo {author} {\bibfnamefont {H.}~\bibnamefont {Moon}}, \bibinfo {author} {\bibfnamefont {L.}~\bibnamefont {Mennel}}, \bibinfo {author} {\bibfnamefont {D.~A.}\ \bibnamefont {Rhodes}}, \bibinfo {author} {\bibfnamefont {H.}~\bibnamefont {Yu}}, \bibinfo {author} {\bibfnamefont {T.}~\bibnamefont {Taniguchi}}, \bibinfo {author} {\bibfnamefont {K.}~\bibnamefont {Watanabe}}, \bibinfo {author} {\bibfnamefont {J.}~\bibnamefont {Yan}}, \bibinfo {author} {\bibfnamefont {D.~G.}\ \bibnamefont {Mandrus}}, \bibinfo {author} {\bibfnamefont {H.}~\bibnamefont {Dery}}, \bibinfo {author} {\bibfnamefont {A.}~\bibnamefont {Pasupathy}}, \bibinfo {author} {\bibfnamefont {D.}~\bibnamefont {Englund}},
  \bibinfo {author} {\bibfnamefont {J.}~\bibnamefont {Hone}}, \bibinfo {author} {\bibfnamefont {W.}~\bibnamefont {Yao}}, \ and\ \bibinfo {author} {\bibfnamefont {X.}~\bibnamefont {Xu}},\ }\href {\doibase 10.1038/s41467-021-21158-8} {\bibfield  {journal} {\bibinfo  {journal} {Nature Communications}\ }\textbf {\bibinfo {volume} {12}} (\bibinfo {year} {2021}),\ 10.1038/s41467-021-21158-8}\BibitemShut {NoStop}%
\bibitem [{\citenamefont {Rosati}\ \emph {et~al.}(2020{\natexlab{a}})\citenamefont {Rosati}, \citenamefont {Wagner}, \citenamefont {Brem}, \citenamefont {Perea-Causin}, \citenamefont {Wietek}, \citenamefont {Zipfel}, \citenamefont {Ziegler}, \citenamefont {Selig}, \citenamefont {Taniguchi}, \citenamefont {Watanabe} \emph {et~al.}}]{rosati2020temporal}%
  \BibitemOpen
  \bibfield  {author} {\bibinfo {author} {\bibfnamefont {R.}~\bibnamefont {Rosati}}, \bibinfo {author} {\bibfnamefont {K.}~\bibnamefont {Wagner}}, \bibinfo {author} {\bibfnamefont {S.}~\bibnamefont {Brem}}, \bibinfo {author} {\bibfnamefont {R.}~\bibnamefont {Perea-Causin}}, \bibinfo {author} {\bibfnamefont {E.}~\bibnamefont {Wietek}}, \bibinfo {author} {\bibfnamefont {J.}~\bibnamefont {Zipfel}}, \bibinfo {author} {\bibfnamefont {J.~D.}\ \bibnamefont {Ziegler}}, \bibinfo {author} {\bibfnamefont {M.}~\bibnamefont {Selig}}, \bibinfo {author} {\bibfnamefont {T.}~\bibnamefont {Taniguchi}}, \bibinfo {author} {\bibfnamefont {K.}~\bibnamefont {Watanabe}},  \emph {et~al.},\ }\href@noop {} {\bibfield  {journal} {\bibinfo  {journal} {ACS Photonics}\ }\textbf {\bibinfo {volume} {7}},\ \bibinfo {pages} {2756} (\bibinfo {year} {2020}{\natexlab{a}})}\BibitemShut {NoStop}%
\bibitem [{\citenamefont {Rosati}\ \emph {et~al.}(2020{\natexlab{b}})\citenamefont {Rosati}, \citenamefont {Brem}, \citenamefont {Perea-Caus{\'\i}n}, \citenamefont {Schmidt}, \citenamefont {Niehues}, \citenamefont {de~Vasconcellos}, \citenamefont {Bratschitsch},\ and\ \citenamefont {Malic}}]{rosati2020strain}%
  \BibitemOpen
  \bibfield  {author} {\bibinfo {author} {\bibfnamefont {R.}~\bibnamefont {Rosati}}, \bibinfo {author} {\bibfnamefont {S.}~\bibnamefont {Brem}}, \bibinfo {author} {\bibfnamefont {R.}~\bibnamefont {Perea-Caus{\'\i}n}}, \bibinfo {author} {\bibfnamefont {R.}~\bibnamefont {Schmidt}}, \bibinfo {author} {\bibfnamefont {I.}~\bibnamefont {Niehues}}, \bibinfo {author} {\bibfnamefont {S.~M.}\ \bibnamefont {de~Vasconcellos}}, \bibinfo {author} {\bibfnamefont {R.}~\bibnamefont {Bratschitsch}}, \ and\ \bibinfo {author} {\bibfnamefont {E.}~\bibnamefont {Malic}},\ }\href@noop {} {\bibfield  {journal} {\bibinfo  {journal} {2D Materials}\ }\textbf {\bibinfo {volume} {8}},\ \bibinfo {pages} {015030} (\bibinfo {year} {2020}{\natexlab{b}})}\BibitemShut {NoStop}%
\bibitem [{\citenamefont {Kirchhof}\ \emph {et~al.}(2022)\citenamefont {Kirchhof}, \citenamefont {Yu}, \citenamefont {Antheaume}, \citenamefont {Gordeev}, \citenamefont {Yagodkin}, \citenamefont {Elliott}, \citenamefont {De~Ara{\'u}jo}, \citenamefont {Sharma}, \citenamefont {Reich},\ and\ \citenamefont {Bolotin}}]{kirchhof2022nanomechanical}%
  \BibitemOpen
  \bibfield  {author} {\bibinfo {author} {\bibfnamefont {J.~N.}\ \bibnamefont {Kirchhof}}, \bibinfo {author} {\bibfnamefont {Y.}~\bibnamefont {Yu}}, \bibinfo {author} {\bibfnamefont {G.}~\bibnamefont {Antheaume}}, \bibinfo {author} {\bibfnamefont {G.}~\bibnamefont {Gordeev}}, \bibinfo {author} {\bibfnamefont {D.}~\bibnamefont {Yagodkin}}, \bibinfo {author} {\bibfnamefont {P.}~\bibnamefont {Elliott}}, \bibinfo {author} {\bibfnamefont {D.~B.}\ \bibnamefont {De~Ara{\'u}jo}}, \bibinfo {author} {\bibfnamefont {S.}~\bibnamefont {Sharma}}, \bibinfo {author} {\bibfnamefont {S.}~\bibnamefont {Reich}}, \ and\ \bibinfo {author} {\bibfnamefont {K.~I.}\ \bibnamefont {Bolotin}},\ }\href@noop {} {\bibfield  {journal} {\bibinfo  {journal} {Nano letters}\ }\textbf {\bibinfo {volume} {22}},\ \bibinfo {pages} {8037} (\bibinfo {year} {2022})}\BibitemShut {NoStop}%
\bibitem [{\citenamefont {Carrascoso}\ \emph {et~al.}(2021)\citenamefont {Carrascoso}, \citenamefont {Li}, \citenamefont {Frisenda},\ and\ \citenamefont {Castellanos-Gomez}}]{36}%
  \BibitemOpen
  \bibfield  {author} {\bibinfo {author} {\bibfnamefont {F.}~\bibnamefont {Carrascoso}}, \bibinfo {author} {\bibfnamefont {H.}~\bibnamefont {Li}}, \bibinfo {author} {\bibfnamefont {R.}~\bibnamefont {Frisenda}}, \ and\ \bibinfo {author} {\bibfnamefont {A.}~\bibnamefont {Castellanos-Gomez}},\ }\href {\doibase 10.1007/s12274-020-2918-2} {\bibfield  {journal} {\bibinfo  {journal} {Nano Research}\ }\textbf {\bibinfo {volume} {14}},\ \bibinfo {pages} {1698} (\bibinfo {year} {2021})}\BibitemShut {NoStop}%
\bibitem [{\citenamefont {Wagner}\ \emph {et~al.}(2020)\citenamefont {Wagner}, \citenamefont {Wietek}, \citenamefont {Ziegler}, \citenamefont {Semina}, \citenamefont {Taniguchi}, \citenamefont {Watanabe}, \citenamefont {Zipfel}, \citenamefont {Glazov},\ and\ \citenamefont {Chernikov}}]{37}%
  \BibitemOpen
  \bibfield  {author} {\bibinfo {author} {\bibfnamefont {K.}~\bibnamefont {Wagner}}, \bibinfo {author} {\bibfnamefont {E.}~\bibnamefont {Wietek}}, \bibinfo {author} {\bibfnamefont {J.~D.}\ \bibnamefont {Ziegler}}, \bibinfo {author} {\bibfnamefont {M.~A.}\ \bibnamefont {Semina}}, \bibinfo {author} {\bibfnamefont {T.}~\bibnamefont {Taniguchi}}, \bibinfo {author} {\bibfnamefont {K.}~\bibnamefont {Watanabe}}, \bibinfo {author} {\bibfnamefont {J.}~\bibnamefont {Zipfel}}, \bibinfo {author} {\bibfnamefont {M.~M.}\ \bibnamefont {Glazov}}, \ and\ \bibinfo {author} {\bibfnamefont {A.}~\bibnamefont {Chernikov}},\ }\href@noop {} {\bibfield  {journal} {\bibinfo  {journal} {Physical review letters}\ }\textbf {\bibinfo {volume} {125}},\ \bibinfo {pages} {267401} (\bibinfo {year} {2020})}\BibitemShut {NoStop}%
\bibitem [{\citenamefont {He}\ \emph {et~al.}(2020)\citenamefont {He}, \citenamefont {Rivera}, \citenamefont {Van~Tuan}, \citenamefont {Wilson}, \citenamefont {Yang}, \citenamefont {Taniguchi}, \citenamefont {Watanabe}, \citenamefont {Yan}, \citenamefont {Mandrus}, \citenamefont {Yu} \emph {et~al.}}]{38}%
  \BibitemOpen
  \bibfield  {author} {\bibinfo {author} {\bibfnamefont {M.}~\bibnamefont {He}}, \bibinfo {author} {\bibfnamefont {P.}~\bibnamefont {Rivera}}, \bibinfo {author} {\bibfnamefont {D.}~\bibnamefont {Van~Tuan}}, \bibinfo {author} {\bibfnamefont {N.~P.}\ \bibnamefont {Wilson}}, \bibinfo {author} {\bibfnamefont {M.}~\bibnamefont {Yang}}, \bibinfo {author} {\bibfnamefont {T.}~\bibnamefont {Taniguchi}}, \bibinfo {author} {\bibfnamefont {K.}~\bibnamefont {Watanabe}}, \bibinfo {author} {\bibfnamefont {J.}~\bibnamefont {Yan}}, \bibinfo {author} {\bibfnamefont {D.~G.}\ \bibnamefont {Mandrus}}, \bibinfo {author} {\bibfnamefont {H.}~\bibnamefont {Yu}},  \emph {et~al.},\ }\href@noop {} {\bibfield  {journal} {\bibinfo  {journal} {Nature communications}\ }\textbf {\bibinfo {volume} {11}},\ \bibinfo {pages} {618} (\bibinfo {year} {2020})}\BibitemShut {NoStop}%
\bibitem [{\citenamefont {Bergh{\"a}user}\ \emph {et~al.}(2018)\citenamefont {Bergh{\"a}user}, \citenamefont {Steinleitner}, \citenamefont {Merkl}, \citenamefont {Huber}, \citenamefont {Knorr},\ and\ \citenamefont {Malic}}]{berghauser2018mapping}%
  \BibitemOpen
  \bibfield  {author} {\bibinfo {author} {\bibfnamefont {G.}~\bibnamefont {Bergh{\"a}user}}, \bibinfo {author} {\bibfnamefont {P.}~\bibnamefont {Steinleitner}}, \bibinfo {author} {\bibfnamefont {P.}~\bibnamefont {Merkl}}, \bibinfo {author} {\bibfnamefont {R.}~\bibnamefont {Huber}}, \bibinfo {author} {\bibfnamefont {A.}~\bibnamefont {Knorr}}, \ and\ \bibinfo {author} {\bibfnamefont {E.}~\bibnamefont {Malic}},\ }\href@noop {} {\bibfield  {journal} {\bibinfo  {journal} {Physical Review B}\ }\textbf {\bibinfo {volume} {98}},\ \bibinfo {pages} {020301} (\bibinfo {year} {2018})}\BibitemShut {NoStop}%
\bibitem [{\citenamefont {Huang}\ \emph {et~al.}(2022)\citenamefont {Huang}, \citenamefont {Zhao}, \citenamefont {Bo}, \citenamefont {Chu}, \citenamefont {Tian}, \citenamefont {Liu}, \citenamefont {Yuan}, \citenamefont {Wu}, \citenamefont {Zhao}, \citenamefont {Xian}, \citenamefont {Watanabe}, \citenamefont {Taniguchi}, \citenamefont {Yang}, \citenamefont {Shi}, \citenamefont {Du}, \citenamefont {Sun}, \citenamefont {Meng}, \citenamefont {Yang},\ and\ \citenamefont {Zhang}}]{41}%
  \BibitemOpen
  \bibfield  {author} {\bibinfo {author} {\bibfnamefont {Z.}~\bibnamefont {Huang}}, \bibinfo {author} {\bibfnamefont {Y.}~\bibnamefont {Zhao}}, \bibinfo {author} {\bibfnamefont {T.}~\bibnamefont {Bo}}, \bibinfo {author} {\bibfnamefont {Y.}~\bibnamefont {Chu}}, \bibinfo {author} {\bibfnamefont {J.}~\bibnamefont {Tian}}, \bibinfo {author} {\bibfnamefont {L.}~\bibnamefont {Liu}}, \bibinfo {author} {\bibfnamefont {Y.}~\bibnamefont {Yuan}}, \bibinfo {author} {\bibfnamefont {F.}~\bibnamefont {Wu}}, \bibinfo {author} {\bibfnamefont {J.}~\bibnamefont {Zhao}}, \bibinfo {author} {\bibfnamefont {L.}~\bibnamefont {Xian}}, \bibinfo {author} {\bibfnamefont {K.}~\bibnamefont {Watanabe}}, \bibinfo {author} {\bibfnamefont {T.}~\bibnamefont {Taniguchi}}, \bibinfo {author} {\bibfnamefont {R.}~\bibnamefont {Yang}}, \bibinfo {author} {\bibfnamefont {D.}~\bibnamefont {Shi}}, \bibinfo {author} {\bibfnamefont {L.}~\bibnamefont {Du}}, \bibinfo {author} {\bibfnamefont {Z.}~\bibnamefont {Sun}}, \bibinfo {author} {\bibfnamefont
  {S.}~\bibnamefont {Meng}}, \bibinfo {author} {\bibfnamefont {W.}~\bibnamefont {Yang}}, \ and\ \bibinfo {author} {\bibfnamefont {G.}~\bibnamefont {Zhang}},\ }\href {\doibase 10.1103/PhysRevB.105.L041409} {\bibfield  {journal} {\bibinfo  {journal} {Physical Review B}\ }\textbf {\bibinfo {volume} {105}} (\bibinfo {year} {2022}),\ 10.1103/PhysRevB.105.L041409}\BibitemShut {NoStop}%
\bibitem [{\citenamefont {Perea-Causin}\ \emph {et~al.}(2023)\citenamefont {Perea-Causin}, \citenamefont {Brem}, \citenamefont {Schmidt},\ and\ \citenamefont {Malic}}]{39}%
  \BibitemOpen
  \bibfield  {author} {\bibinfo {author} {\bibfnamefont {R.}~\bibnamefont {Perea-Causin}}, \bibinfo {author} {\bibfnamefont {S.}~\bibnamefont {Brem}}, \bibinfo {author} {\bibfnamefont {O.}~\bibnamefont {Schmidt}}, \ and\ \bibinfo {author} {\bibfnamefont {E.}~\bibnamefont {Malic}},\ }\href {http://arxiv.org/abs/2306.10812} {\enquote {\bibinfo {title} {Trion photoluminescence and trion stability in atomically thin semiconductors},}\ } (\bibinfo {year} {2023})\BibitemShut {NoStop}%
\bibitem [{\citenamefont {Brem}\ \emph {et~al.}(2020)\citenamefont {Brem}, \citenamefont {Ekman}, \citenamefont {Christiansen}, \citenamefont {Katsch}, \citenamefont {Selig}, \citenamefont {Robert}, \citenamefont {Marie}, \citenamefont {Urbaszek}, \citenamefont {Knorr},\ and\ \citenamefont {Malic}}]{40}%
  \BibitemOpen
  \bibfield  {author} {\bibinfo {author} {\bibfnamefont {S.}~\bibnamefont {Brem}}, \bibinfo {author} {\bibfnamefont {A.}~\bibnamefont {Ekman}}, \bibinfo {author} {\bibfnamefont {D.}~\bibnamefont {Christiansen}}, \bibinfo {author} {\bibfnamefont {F.}~\bibnamefont {Katsch}}, \bibinfo {author} {\bibfnamefont {M.}~\bibnamefont {Selig}}, \bibinfo {author} {\bibfnamefont {C.}~\bibnamefont {Robert}}, \bibinfo {author} {\bibfnamefont {X.}~\bibnamefont {Marie}}, \bibinfo {author} {\bibfnamefont {B.}~\bibnamefont {Urbaszek}}, \bibinfo {author} {\bibfnamefont {A.}~\bibnamefont {Knorr}}, \ and\ \bibinfo {author} {\bibfnamefont {E.}~\bibnamefont {Malic}},\ }\href {\doibase 10.1021/acs.nanolett.0c00633} {\bibfield  {journal} {\bibinfo  {journal} {Nano Letters}\ }\textbf {\bibinfo {volume} {20}} (\bibinfo {year} {2020}),\ 10.1021/acs.nanolett.0c00633}\BibitemShut {NoStop}%
\bibitem [{\citenamefont {Klots}\ \emph {et~al.}(2018)\citenamefont {Klots}, \citenamefont {Weintrub}, \citenamefont {Prasai}, \citenamefont {Kidd}, \citenamefont {Varga}, \citenamefont {Velizhanin},\ and\ \citenamefont {Bolotin}}]{klots2018controlled}%
  \BibitemOpen
  \bibfield  {author} {\bibinfo {author} {\bibfnamefont {A.~R.}\ \bibnamefont {Klots}}, \bibinfo {author} {\bibfnamefont {B.}~\bibnamefont {Weintrub}}, \bibinfo {author} {\bibfnamefont {D.}~\bibnamefont {Prasai}}, \bibinfo {author} {\bibfnamefont {D.}~\bibnamefont {Kidd}}, \bibinfo {author} {\bibfnamefont {K.}~\bibnamefont {Varga}}, \bibinfo {author} {\bibfnamefont {K.~A.}\ \bibnamefont {Velizhanin}}, \ and\ \bibinfo {author} {\bibfnamefont {K.~I.}\ \bibnamefont {Bolotin}},\ }\href@noop {} {\bibfield  {journal} {\bibinfo  {journal} {Scientific reports}\ }\textbf {\bibinfo {volume} {8}},\ \bibinfo {pages} {768} (\bibinfo {year} {2018})}\BibitemShut {NoStop}%
\bibitem [{\citenamefont {Chakraborty}\ \emph {et~al.}(2015)\citenamefont {Chakraborty}, \citenamefont {Kinnischtzke}, \citenamefont {Goodfellow}, \citenamefont {Beams},\ and\ \citenamefont {Vamivakas}}]{42}%
  \BibitemOpen
  \bibfield  {author} {\bibinfo {author} {\bibfnamefont {C.}~\bibnamefont {Chakraborty}}, \bibinfo {author} {\bibfnamefont {L.}~\bibnamefont {Kinnischtzke}}, \bibinfo {author} {\bibfnamefont {K.~M.}\ \bibnamefont {Goodfellow}}, \bibinfo {author} {\bibfnamefont {R.}~\bibnamefont {Beams}}, \ and\ \bibinfo {author} {\bibfnamefont {A.~N.}\ \bibnamefont {Vamivakas}},\ }\href {\doibase 10.1038/nnano.2015.79} {\ \textbf {\bibinfo {volume} {10}},\ \bibinfo {pages} {507} (\bibinfo {year} {2015})}\BibitemShut {NoStop}%
\bibitem [{43(2015)}]{43}%
  \BibitemOpen
  \href {\doibase 10.1038/nnano.2015.75} {\ \textbf {\bibinfo {volume} {10}},\ \bibinfo {pages} {497} (\bibinfo {year} {2015})}\BibitemShut {NoStop}%
\bibitem [{\citenamefont {Srivastava}\ \emph {et~al.}(2015)\citenamefont {Srivastava}, \citenamefont {Sidler}, \citenamefont {Allain}, \citenamefont {Lembke}, \citenamefont {Kis},\ and\ \citenamefont {Imamoglu}}]{44}%
  \BibitemOpen
  \bibfield  {author} {\bibinfo {author} {\bibfnamefont {A.}~\bibnamefont {Srivastava}}, \bibinfo {author} {\bibfnamefont {M.}~\bibnamefont {Sidler}}, \bibinfo {author} {\bibfnamefont {A.~V.}\ \bibnamefont {Allain}}, \bibinfo {author} {\bibfnamefont {D.~S.}\ \bibnamefont {Lembke}}, \bibinfo {author} {\bibfnamefont {A.}~\bibnamefont {Kis}}, \ and\ \bibinfo {author} {\bibfnamefont {A.}~\bibnamefont {Imamoglu}},\ }\href {\doibase 10.1038/nnano.2015.60} {\bibfield  {journal} {\bibinfo  {journal} {Nature Nanotechnology}\ }\textbf {\bibinfo {volume} {10}} (\bibinfo {year} {2015}),\ 10.1038/nnano.2015.60}\BibitemShut {NoStop}%
\bibitem [{\citenamefont {Koperski}\ \emph {et~al.}(2015)\citenamefont {Koperski}, \citenamefont {Nogajewski}, \citenamefont {Arora}, \citenamefont {Cherkez}, \citenamefont {Mallet}, \citenamefont {Veuillen}, \citenamefont {Marcus}, \citenamefont {Kossacki},\ and\ \citenamefont {Potemski}}]{45}%
  \BibitemOpen
  \bibfield  {author} {\bibinfo {author} {\bibfnamefont {M.}~\bibnamefont {Koperski}}, \bibinfo {author} {\bibfnamefont {K.}~\bibnamefont {Nogajewski}}, \bibinfo {author} {\bibfnamefont {A.}~\bibnamefont {Arora}}, \bibinfo {author} {\bibfnamefont {V.}~\bibnamefont {Cherkez}}, \bibinfo {author} {\bibfnamefont {P.}~\bibnamefont {Mallet}}, \bibinfo {author} {\bibfnamefont {J.-Y.}\ \bibnamefont {Veuillen}}, \bibinfo {author} {\bibfnamefont {J.}~\bibnamefont {Marcus}}, \bibinfo {author} {\bibfnamefont {P.}~\bibnamefont {Kossacki}}, \ and\ \bibinfo {author} {\bibfnamefont {M.}~\bibnamefont {Potemski}},\ }\href {\doibase 10.1038/nnano.2015.67} {\bibfield  {journal} {\bibinfo  {journal} {Nature Nanotechnology}\ }\textbf {\bibinfo {volume} {10}} (\bibinfo {year} {2015}),\ 10.1038/nnano.2015.67}\BibitemShut {NoStop}%
\bibitem [{\citenamefont {Khan}\ \emph {et~al.}(2020)\citenamefont {Khan}, \citenamefont {Lu}, \citenamefont {Ma}, \citenamefont {Lu},\ and\ \citenamefont {Liu}}]{46}%
  \BibitemOpen
  \bibfield  {author} {\bibinfo {author} {\bibfnamefont {A.~R.}\ \bibnamefont {Khan}}, \bibinfo {author} {\bibfnamefont {T.}~\bibnamefont {Lu}}, \bibinfo {author} {\bibfnamefont {W.}~\bibnamefont {Ma}}, \bibinfo {author} {\bibfnamefont {Y.}~\bibnamefont {Lu}}, \ and\ \bibinfo {author} {\bibfnamefont {Y.}~\bibnamefont {Liu}},\ }\href {\doibase 10.1002/aelm.201901381} {\bibfield  {journal} {\bibinfo  {journal} {Advanced Electronic Materials}\ }\textbf {\bibinfo {volume} {6}},\ \bibinfo {pages} {1901381} (\bibinfo {year} {2020})}\BibitemShut {NoStop}%
\bibitem [{\citenamefont {Castellanos-Gomez}\ \emph {et~al.}(2013)\citenamefont {Castellanos-Gomez}, \citenamefont {Roldán}, \citenamefont {Cappelluti}, \citenamefont {Buscema}, \citenamefont {Guinea}, \citenamefont {Van Der~Zant},\ and\ \citenamefont {Steele}}]{47}%
  \BibitemOpen
  \bibfield  {author} {\bibinfo {author} {\bibfnamefont {A.}~\bibnamefont {Castellanos-Gomez}}, \bibinfo {author} {\bibfnamefont {R.}~\bibnamefont {Roldán}}, \bibinfo {author} {\bibfnamefont {E.}~\bibnamefont {Cappelluti}}, \bibinfo {author} {\bibfnamefont {M.}~\bibnamefont {Buscema}}, \bibinfo {author} {\bibfnamefont {F.}~\bibnamefont {Guinea}}, \bibinfo {author} {\bibfnamefont {H.~S.}\ \bibnamefont {Van Der~Zant}}, \ and\ \bibinfo {author} {\bibfnamefont {G.~A.}\ \bibnamefont {Steele}},\ }\href {\doibase 10.1021/nl402875m} {\bibfield  {journal} {\bibinfo  {journal} {Nano Letters}\ }\textbf {\bibinfo {volume} {13}},\ \bibinfo {pages} {5361} (\bibinfo {year} {2013})}\BibitemShut {NoStop}%
\bibitem [{\citenamefont {Toth}\ and\ \citenamefont {Aharonovich}(2019)}]{48}%
  \BibitemOpen
  \bibfield  {author} {\bibinfo {author} {\bibfnamefont {M.}~\bibnamefont {Toth}}\ and\ \bibinfo {author} {\bibfnamefont {I.}~\bibnamefont {Aharonovich}},\ }\href {\doibase 10.1146/annurev-physchem-042018-052628} {\bibfield  {journal} {\bibinfo  {journal} {Annual Review of Physical Chemistry}\ }\textbf {\bibinfo {volume} {70}},\ \bibinfo {pages} {123} (\bibinfo {year} {2019})}\BibitemShut {NoStop}%
\bibitem [{\citenamefont {Iff}\ \emph {et~al.}(2019)\citenamefont {Iff}, \citenamefont {Tedeschi}, \citenamefont {Martín-Sánchez}, \citenamefont {Moczała-Dusanowska}, \citenamefont {Tongay}, \citenamefont {Yumigeta}, \citenamefont {Taboada-Gutiérrez}, \citenamefont {Savaresi}, \citenamefont {Rastelli}, \citenamefont {Alonso-González}, \citenamefont {Höfling}, \citenamefont {Trotta},\ and\ \citenamefont {Schneider}}]{49}%
  \BibitemOpen
  \bibfield  {author} {\bibinfo {author} {\bibfnamefont {O.}~\bibnamefont {Iff}}, \bibinfo {author} {\bibfnamefont {D.}~\bibnamefont {Tedeschi}}, \bibinfo {author} {\bibfnamefont {J.}~\bibnamefont {Martín-Sánchez}}, \bibinfo {author} {\bibfnamefont {M.}~\bibnamefont {Moczała-Dusanowska}}, \bibinfo {author} {\bibfnamefont {S.}~\bibnamefont {Tongay}}, \bibinfo {author} {\bibfnamefont {K.}~\bibnamefont {Yumigeta}}, \bibinfo {author} {\bibfnamefont {J.}~\bibnamefont {Taboada-Gutiérrez}}, \bibinfo {author} {\bibfnamefont {M.}~\bibnamefont {Savaresi}}, \bibinfo {author} {\bibfnamefont {A.}~\bibnamefont {Rastelli}}, \bibinfo {author} {\bibfnamefont {P.}~\bibnamefont {Alonso-González}}, \bibinfo {author} {\bibfnamefont {S.}~\bibnamefont {Höfling}}, \bibinfo {author} {\bibfnamefont {R.}~\bibnamefont {Trotta}}, \ and\ \bibinfo {author} {\bibfnamefont {C.}~\bibnamefont {Schneider}},\ }\href {\doibase 10.1021/acs.nanolett.9b02221} {\bibfield  {journal} {\bibinfo  {journal} {Nano Letters}\ }\textbf {\bibinfo
  {volume} {19}} (\bibinfo {year} {2019}),\ 10.1021/acs.nanolett.9b02221}\BibitemShut {NoStop}%
\bibitem [{\citenamefont {Savaresi}\ \emph {et~al.}(2023)\citenamefont {Savaresi}, \citenamefont {Martínez-Suárez}, \citenamefont {Tedeschi}, \citenamefont {Ronco}, \citenamefont {Hierro-Rodríguez}, \citenamefont {McVitie}, \citenamefont {Stroj}, \citenamefont {Aberl}, \citenamefont {Brehm}, \citenamefont {García-Suárez}, \citenamefont {Rota}, \citenamefont {Alonso-González}, \citenamefont {Martín-Sánchez},\ and\ \citenamefont {Trotta}}]{50}%
  \BibitemOpen
  \bibfield  {author} {\bibinfo {author} {\bibfnamefont {M.}~\bibnamefont {Savaresi}}, \bibinfo {author} {\bibfnamefont {A.}~\bibnamefont {Martínez-Suárez}}, \bibinfo {author} {\bibfnamefont {D.}~\bibnamefont {Tedeschi}}, \bibinfo {author} {\bibfnamefont {G.}~\bibnamefont {Ronco}}, \bibinfo {author} {\bibfnamefont {A.}~\bibnamefont {Hierro-Rodríguez}}, \bibinfo {author} {\bibfnamefont {S.}~\bibnamefont {McVitie}}, \bibinfo {author} {\bibfnamefont {S.}~\bibnamefont {Stroj}}, \bibinfo {author} {\bibfnamefont {J.}~\bibnamefont {Aberl}}, \bibinfo {author} {\bibfnamefont {M.}~\bibnamefont {Brehm}}, \bibinfo {author} {\bibfnamefont {V.~M.}\ \bibnamefont {García-Suárez}}, \bibinfo {author} {\bibfnamefont {M.~B.}\ \bibnamefont {Rota}}, \bibinfo {author} {\bibfnamefont {P.}~\bibnamefont {Alonso-González}}, \bibinfo {author} {\bibfnamefont {J.}~\bibnamefont {Martín-Sánchez}}, \ and\ \bibinfo {author} {\bibfnamefont {R.}~\bibnamefont {Trotta}},\ }\href {\doibase 10.48550/arXiv.2301.10273} {\enquote {\bibinfo
  {title} {Strain-induced dynamic control over the population of quantum emitters in two-dimensional materials},}\ } (\bibinfo {year} {2023})\BibitemShut {NoStop}%
\bibitem [{\citenamefont {Patel}\ \emph {et~al.}(2010)\citenamefont {Patel}, \citenamefont {Bennett}, \citenamefont {Farrer}, \citenamefont {Nicoll}, \citenamefont {Ritchie},\ and\ \citenamefont {Shields}}]{patel2010two}%
  \BibitemOpen
  \bibfield  {author} {\bibinfo {author} {\bibfnamefont {R.~B.}\ \bibnamefont {Patel}}, \bibinfo {author} {\bibfnamefont {A.~J.}\ \bibnamefont {Bennett}}, \bibinfo {author} {\bibfnamefont {I.}~\bibnamefont {Farrer}}, \bibinfo {author} {\bibfnamefont {C.~A.}\ \bibnamefont {Nicoll}}, \bibinfo {author} {\bibfnamefont {D.~A.}\ \bibnamefont {Ritchie}}, \ and\ \bibinfo {author} {\bibfnamefont {A.~J.}\ \bibnamefont {Shields}},\ }\href@noop {} {\bibfield  {journal} {\bibinfo  {journal} {Nature photonics}\ }\textbf {\bibinfo {volume} {4}},\ \bibinfo {pages} {632} (\bibinfo {year} {2010})}\BibitemShut {NoStop}%
\bibitem [{\citenamefont {Barr{\'e}}\ \emph {et~al.}(2022)\citenamefont {Barr{\'e}}, \citenamefont {Karni}, \citenamefont {Liu}, \citenamefont {O’Beirne}, \citenamefont {Chen}, \citenamefont {Ribeiro}, \citenamefont {Yu}, \citenamefont {Kim}, \citenamefont {Watanabe}, \citenamefont {Taniguchi} \emph {et~al.}}]{51}%
  \BibitemOpen
  \bibfield  {author} {\bibinfo {author} {\bibfnamefont {E.}~\bibnamefont {Barr{\'e}}}, \bibinfo {author} {\bibfnamefont {O.}~\bibnamefont {Karni}}, \bibinfo {author} {\bibfnamefont {E.}~\bibnamefont {Liu}}, \bibinfo {author} {\bibfnamefont {A.~L.}\ \bibnamefont {O’Beirne}}, \bibinfo {author} {\bibfnamefont {X.}~\bibnamefont {Chen}}, \bibinfo {author} {\bibfnamefont {H.~B.}\ \bibnamefont {Ribeiro}}, \bibinfo {author} {\bibfnamefont {L.}~\bibnamefont {Yu}}, \bibinfo {author} {\bibfnamefont {B.}~\bibnamefont {Kim}}, \bibinfo {author} {\bibfnamefont {K.}~\bibnamefont {Watanabe}}, \bibinfo {author} {\bibfnamefont {T.}~\bibnamefont {Taniguchi}},  \emph {et~al.},\ }\href@noop {} {\bibfield  {journal} {\bibinfo  {journal} {Science}\ }\textbf {\bibinfo {volume} {376}},\ \bibinfo {pages} {406} (\bibinfo {year} {2022})}\BibitemShut {NoStop}%
\bibitem [{\citenamefont {Dey}\ \emph {et~al.}(2020)\citenamefont {Dey}, \citenamefont {Richter}, \citenamefont {Debnath}, \citenamefont {Huang}, \citenamefont {Polavarapu},\ and\ \citenamefont {Feldmann}}]{52}%
  \BibitemOpen
  \bibfield  {author} {\bibinfo {author} {\bibfnamefont {A.}~\bibnamefont {Dey}}, \bibinfo {author} {\bibfnamefont {A.~F.}\ \bibnamefont {Richter}}, \bibinfo {author} {\bibfnamefont {T.}~\bibnamefont {Debnath}}, \bibinfo {author} {\bibfnamefont {H.}~\bibnamefont {Huang}}, \bibinfo {author} {\bibfnamefont {L.}~\bibnamefont {Polavarapu}}, \ and\ \bibinfo {author} {\bibfnamefont {J.}~\bibnamefont {Feldmann}},\ }\href@noop {} {\bibfield  {journal} {\bibinfo  {journal} {ACS nano}\ }\textbf {\bibinfo {volume} {14}},\ \bibinfo {pages} {5855} (\bibinfo {year} {2020})}\BibitemShut {NoStop}%
\bibitem [{\citenamefont {Zhang}\ \emph {et~al.}(2017)\citenamefont {Zhang}, \citenamefont {Yin}, \citenamefont {Parida}, \citenamefont {Ahmed}, \citenamefont {Pan}, \citenamefont {Bakr}, \citenamefont {Brédas},\ and\ \citenamefont {Mohammed}}]{53}%
  \BibitemOpen
  \bibfield  {author} {\bibinfo {author} {\bibfnamefont {Y.}~\bibnamefont {Zhang}}, \bibinfo {author} {\bibfnamefont {J.}~\bibnamefont {Yin}}, \bibinfo {author} {\bibfnamefont {M.~R.}\ \bibnamefont {Parida}}, \bibinfo {author} {\bibfnamefont {G.~H.}\ \bibnamefont {Ahmed}}, \bibinfo {author} {\bibfnamefont {J.}~\bibnamefont {Pan}}, \bibinfo {author} {\bibfnamefont {O.~M.}\ \bibnamefont {Bakr}}, \bibinfo {author} {\bibfnamefont {J.-L.}\ \bibnamefont {Brédas}}, \ and\ \bibinfo {author} {\bibfnamefont {O.~F.}\ \bibnamefont {Mohammed}},\ }\href {\doibase 10.1021/acs.jpclett.7b01381} {\bibfield  {journal} {\bibinfo  {journal} {The Journal of Physical Chemistry Letters}\ }\textbf {\bibinfo {volume} {8}} (\bibinfo {year} {2017}),\ 10.1021/acs.jpclett.7b01381}\BibitemShut {NoStop}%
\bibitem [{\citenamefont {Cenker}\ \emph {et~al.}(2022)\citenamefont {Cenker}, \citenamefont {Sivakumar}, \citenamefont {Xie}, \citenamefont {Miller}, \citenamefont {Thijssen}, \citenamefont {Liu}, \citenamefont {Dismukes}, \citenamefont {Fonseca}, \citenamefont {Anderson}, \citenamefont {Zhu}, \citenamefont {Roy}, \citenamefont {Xiao}, \citenamefont {Chu}, \citenamefont {Cao},\ and\ \citenamefont {Xu}}]{Cenker2022}%
  \BibitemOpen
  \bibfield  {author} {\bibinfo {author} {\bibfnamefont {J.}~\bibnamefont {Cenker}}, \bibinfo {author} {\bibfnamefont {S.}~\bibnamefont {Sivakumar}}, \bibinfo {author} {\bibfnamefont {K.}~\bibnamefont {Xie}}, \bibinfo {author} {\bibfnamefont {A.}~\bibnamefont {Miller}}, \bibinfo {author} {\bibfnamefont {P.}~\bibnamefont {Thijssen}}, \bibinfo {author} {\bibfnamefont {Z.}~\bibnamefont {Liu}}, \bibinfo {author} {\bibfnamefont {A.}~\bibnamefont {Dismukes}}, \bibinfo {author} {\bibfnamefont {J.}~\bibnamefont {Fonseca}}, \bibinfo {author} {\bibfnamefont {E.}~\bibnamefont {Anderson}}, \bibinfo {author} {\bibfnamefont {X.}~\bibnamefont {Zhu}}, \bibinfo {author} {\bibfnamefont {X.}~\bibnamefont {Roy}}, \bibinfo {author} {\bibfnamefont {D.}~\bibnamefont {Xiao}}, \bibinfo {author} {\bibfnamefont {J.-H.}\ \bibnamefont {Chu}}, \bibinfo {author} {\bibfnamefont {T.}~\bibnamefont {Cao}}, \ and\ \bibinfo {author} {\bibfnamefont {X.}~\bibnamefont {Xu}},\ }\href {\doibase 10.1038/s41565-021-01052-6} {\bibfield  {journal}
  {\bibinfo  {journal} {Nature Nanotechnology}\ }\textbf {\bibinfo {volume} {17}},\ \bibinfo {pages} {256–261} (\bibinfo {year} {2022})}\BibitemShut {NoStop}%
\bibitem [{\citenamefont {Diederich}\ \emph {et~al.}(2022)\citenamefont {Diederich}, \citenamefont {Cenker}, \citenamefont {Ren}, \citenamefont {Fonseca}, \citenamefont {Chica}, \citenamefont {Bae}, \citenamefont {Zhu}, \citenamefont {Roy}, \citenamefont {Cao}, \citenamefont {Xiao},\ and\ \citenamefont {Xu}}]{Diederich2022}%
  \BibitemOpen
  \bibfield  {author} {\bibinfo {author} {\bibfnamefont {G.~M.}\ \bibnamefont {Diederich}}, \bibinfo {author} {\bibfnamefont {J.}~\bibnamefont {Cenker}}, \bibinfo {author} {\bibfnamefont {Y.}~\bibnamefont {Ren}}, \bibinfo {author} {\bibfnamefont {J.}~\bibnamefont {Fonseca}}, \bibinfo {author} {\bibfnamefont {D.~G.}\ \bibnamefont {Chica}}, \bibinfo {author} {\bibfnamefont {Y.~J.}\ \bibnamefont {Bae}}, \bibinfo {author} {\bibfnamefont {X.}~\bibnamefont {Zhu}}, \bibinfo {author} {\bibfnamefont {X.}~\bibnamefont {Roy}}, \bibinfo {author} {\bibfnamefont {T.}~\bibnamefont {Cao}}, \bibinfo {author} {\bibfnamefont {D.}~\bibnamefont {Xiao}}, \ and\ \bibinfo {author} {\bibfnamefont {X.}~\bibnamefont {Xu}},\ }\href {\doibase 10.1038/s41565-022-01259-1} {\bibfield  {journal} {\bibinfo  {journal} {Nature Nanotechnology}\ }\textbf {\bibinfo {volume} {18}},\ \bibinfo {pages} {23–28} (\bibinfo {year} {2022})}\BibitemShut {NoStop}%
\bibitem [{\citenamefont {Beaulieu}\ \emph {et~al.}(2023)\citenamefont {Beaulieu}, \citenamefont {Dong}, \citenamefont {Christiansson}, \citenamefont {Werner}, \citenamefont {Pincelli}, \citenamefont {Ziegler}, \citenamefont {Taniguchi}, \citenamefont {Watanabe}, \citenamefont {Chernikov}, \citenamefont {Wolf}, \citenamefont {Rettig}, \citenamefont {Ernstorfer},\ and\ \citenamefont {Schüler}}]{54}%
  \BibitemOpen
  \bibfield  {author} {\bibinfo {author} {\bibfnamefont {S.}~\bibnamefont {Beaulieu}}, \bibinfo {author} {\bibfnamefont {S.}~\bibnamefont {Dong}}, \bibinfo {author} {\bibfnamefont {V.}~\bibnamefont {Christiansson}}, \bibinfo {author} {\bibfnamefont {P.}~\bibnamefont {Werner}}, \bibinfo {author} {\bibfnamefont {T.}~\bibnamefont {Pincelli}}, \bibinfo {author} {\bibfnamefont {J.~D.}\ \bibnamefont {Ziegler}}, \bibinfo {author} {\bibfnamefont {T.}~\bibnamefont {Taniguchi}}, \bibinfo {author} {\bibfnamefont {K.}~\bibnamefont {Watanabe}}, \bibinfo {author} {\bibfnamefont {A.}~\bibnamefont {Chernikov}}, \bibinfo {author} {\bibfnamefont {M.}~\bibnamefont {Wolf}}, \bibinfo {author} {\bibfnamefont {L.}~\bibnamefont {Rettig}}, \bibinfo {author} {\bibfnamefont {R.}~\bibnamefont {Ernstorfer}}, \ and\ \bibinfo {author} {\bibfnamefont {M.}~\bibnamefont {Schüler}},\ }\href {http://arxiv.org/abs/2308.09634} {\enquote {\bibinfo {title} {Berry {Curvature} {Signatures} in {Chiroptical} {Excitonic} {Transitions}},}\ } (\bibinfo
  {year} {2023})\BibitemShut {NoStop}%
\bibitem [{\citenamefont {Kumar}\ \emph {et~al.}(2021)\citenamefont {Kumar}, \citenamefont {Yagodkin}, \citenamefont {Stetzuhn}, \citenamefont {Kovalchuk}, \citenamefont {Melnikov}, \citenamefont {Elliott}, \citenamefont {Sharma}, \citenamefont {Gahl},\ and\ \citenamefont {Bolotin}}]{55}%
  \BibitemOpen
  \bibfield  {author} {\bibinfo {author} {\bibfnamefont {A.}~\bibnamefont {Kumar}}, \bibinfo {author} {\bibfnamefont {D.}~\bibnamefont {Yagodkin}}, \bibinfo {author} {\bibfnamefont {N.}~\bibnamefont {Stetzuhn}}, \bibinfo {author} {\bibfnamefont {S.}~\bibnamefont {Kovalchuk}}, \bibinfo {author} {\bibfnamefont {A.}~\bibnamefont {Melnikov}}, \bibinfo {author} {\bibfnamefont {P.}~\bibnamefont {Elliott}}, \bibinfo {author} {\bibfnamefont {S.}~\bibnamefont {Sharma}}, \bibinfo {author} {\bibfnamefont {C.}~\bibnamefont {Gahl}}, \ and\ \bibinfo {author} {\bibfnamefont {K.~I.}\ \bibnamefont {Bolotin}},\ }\href {\doibase 10.1021/acs.nanolett.1c01538} {\bibfield  {journal} {\bibinfo  {journal} {Nano Letters}\ }\textbf {\bibinfo {volume} {21}},\ \bibinfo {pages} {7123} (\bibinfo {year} {2021})}\BibitemShut {NoStop}%
\bibitem [{\citenamefont {Yu}\ \emph {et~al.}(2014)\citenamefont {Yu}, \citenamefont {Liu}, \citenamefont {Gong}, \citenamefont {Xu},\ and\ \citenamefont {Yao}}]{56}%
  \BibitemOpen
  \bibfield  {author} {\bibinfo {author} {\bibfnamefont {H.}~\bibnamefont {Yu}}, \bibinfo {author} {\bibfnamefont {G.~B.}\ \bibnamefont {Liu}}, \bibinfo {author} {\bibfnamefont {P.}~\bibnamefont {Gong}}, \bibinfo {author} {\bibfnamefont {X.}~\bibnamefont {Xu}}, \ and\ \bibinfo {author} {\bibfnamefont {W.}~\bibnamefont {Yao}},\ }\href {\doibase 10.1038/ncomms4876} {\bibfield  {journal} {\bibinfo  {journal} {Nature Communications}\ }\textbf {\bibinfo {volume} {5}} (\bibinfo {year} {2014}),\ 10.1038/ncomms4876}\BibitemShut {NoStop}%
\bibitem [{\citenamefont {Glazov}\ \emph {et~al.}(2022)\citenamefont {Glazov}, \citenamefont {Dirnberger}, \citenamefont {Menon}, \citenamefont {Taniguchi}, \citenamefont {Watanabe}, \citenamefont {Bougeard}, \citenamefont {Ziegler},\ and\ \citenamefont {Chernikov}}]{57}%
  \BibitemOpen
  \bibfield  {author} {\bibinfo {author} {\bibfnamefont {M.~M.}\ \bibnamefont {Glazov}}, \bibinfo {author} {\bibfnamefont {F.}~\bibnamefont {Dirnberger}}, \bibinfo {author} {\bibfnamefont {V.~M.}\ \bibnamefont {Menon}}, \bibinfo {author} {\bibfnamefont {T.}~\bibnamefont {Taniguchi}}, \bibinfo {author} {\bibfnamefont {K.}~\bibnamefont {Watanabe}}, \bibinfo {author} {\bibfnamefont {D.}~\bibnamefont {Bougeard}}, \bibinfo {author} {\bibfnamefont {J.~D.}\ \bibnamefont {Ziegler}}, \ and\ \bibinfo {author} {\bibfnamefont {A.}~\bibnamefont {Chernikov}},\ }\href {\doibase 10.1103/PhysRevB.106.125303} {\bibfield  {journal} {\bibinfo  {journal} {Physical Review B}\ }\textbf {\bibinfo {volume} {106}} (\bibinfo {year} {2022}),\ 10.1103/PhysRevB.106.125303}\BibitemShut {NoStop}%
\bibitem [{\citenamefont {Iakovlev}\ and\ \citenamefont {Glazov}(2023)}]{iakovlev2023fermi}%
  \BibitemOpen
  \bibfield  {author} {\bibinfo {author} {\bibfnamefont {Z.}~\bibnamefont {Iakovlev}}\ and\ \bibinfo {author} {\bibfnamefont {M.~M.}\ \bibnamefont {Glazov}},\ }\href@noop {} {\bibfield  {journal} {\bibinfo  {journal} {2D Materials}\ } (\bibinfo {year} {2023})}\BibitemShut {NoStop}%
\end{thebibliography}%

\end{document}